\pdfoutput=1
\documentclass[english]{article}
\usepackage[T1]{fontenc}
\usepackage[utf8]{luainputenc}
\usepackage{geometry}
\usepackage{color}
\usepackage{amsmath}
\usepackage{amssymb}
\usepackage{graphicx}
\usepackage{esint}
\usepackage[numbers]{natbib}

\makeatletter
\newcommand{\lyxaddress}[1]{
\par {\raggedright #1
\vspace{1.4em}
\noindent\par}
}

\@ifundefined{showcaptionsetup}{}{%
 \PassOptionsToPackage{caption=false}{subfig}}
\usepackage{subfig}
\makeatother

\usepackage{babel}
\begin{document}

\title{Current through a multi-lead junction caused by applied bias with
arbitrary time-dependence}

\author{Michael Ridley$^{1,2}$, Angus MacKinnon$^{2}$, Lev Kantorovich$^{1}$}

\maketitle

\lyxaddress{$^{1}$Department of Physics, King's College London, Strand, London,
WC2R 2LS, United Kingdom\\$^{2}$Department of Physics, The Blackett
Laboratory, Imperial College London, South Kensington Campus, London
SW7 2AZ, United Kingdom}
\begin{abstract}
We apply the Nonequilibrium Green's Function (NEGF) formalism to the
problem of a multi-terminal nanojunction subject to an arbitrary time-dependent
bias. In particular, we show that taking a generic one-particle system
Hamiltonian within the wide band limit approximation (WBLA), it is
possible to obtain a closed analytical expression for the current
in each lead. Our formula reduces to the well-known result of Jauho
\textit{et. al.} {[}doi:10.1103/PhysRevB.50.5528{]} in the limit where
the switch-on time is taken to the remote past, and to the result
of Tuovinen \textit{et. al.} {[}doi:10.1088/1742-6596/427/1/012014{]}
when the bias is maintained at a constant value after the switch-on.
As we use a partition-free approach, our formula contains both the
long-time current and transient effects due to the sudden switch-on
of the bias. Numerical calculations performed for the simple case
of a single-level quantum dot coupled to two leads are performed for
a sinusoidally-varying bias. At certain frequencies of the driving
bias, we observe `ringing' oscillations of the current, whose dependence
on the dot level, level width, oscillation amplitude and temperature
is also investigated.
\end{abstract}

\section{Introduction}

The problem of computing the nonlinear current response to a bias
dropped across a nanoscale structure can be treated within the framework
of the Nonequilibrium Green's Function (NEGF) formalism\citep{matsubara-PTP-1955,Martin-Schwinger-PR-1959,konstantinov-perel-JETP-1961,KB1962,Keldysh1964,Langreth1976,RammerSmith1986,RVLS2013}.
This is an exact method for the calculation of time-dependent ensemble
averages in many-electron systems both in and out of equilibrium,
which automatically preserves all required conservation laws\citep{KB1961,RVLS2013}.
The mathematical equivalence of the Green's function approaches to
the nonequilibrium and equilibrium problems has been well understood
since the work of Martin, Schwinger, Matsubara, Keldysh, Kadanoff
and Baym \citep{Martin-Schwinger-PR-1959,matsubara-PTP-1955,Keldysh1964,KB1962}.
Perhaps the most elegant depiction of this equivalence comes with
the introduction of the Konstantinov-Perel' contour, on which the
initial preparation of the system together with its subsequent time
evolution can be conveniently represented\citep{konstantinov-perel-JETP-1961,Keldysh1964}. 

\begin{figure}
\includegraphics[height=5cm]{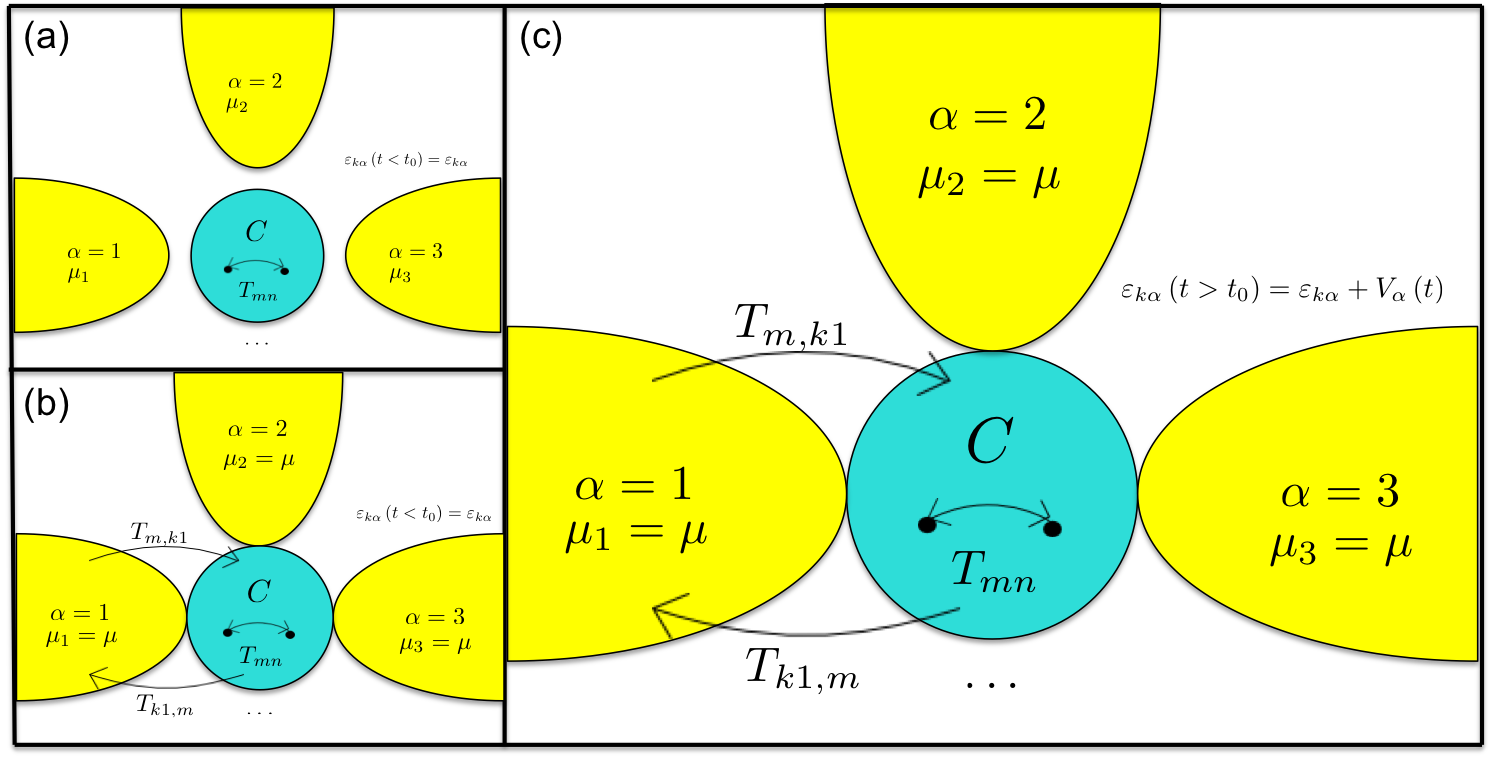}

\caption{Schematic of the multi-terminal nanojunction in the time regimes before
and after $t_{0}$: (a) $t<t_{0}$ in the partitioned approach; (b)
$t<t_{0}$ in the partition-free approach; (c) $t\geq t_{0}$ in either
approach.\label{fig:Nanojunc}}
\end{figure}

In this paper, we are concerned with the problem of electron transport
through a nanojunction, consisting of a small central region $C$,
contacted with a set of conducting leads labelled by $\alpha$, themselves
in contact with an external circuit and battery. This may for example
describe a molecule or quantum dot contacted with a pair of capacitor
plates, as found for example in a single electron transistor\citep{QD1997};
a schematic of the generic set-up is shown in Fig. \ref{fig:Nanojunc}. 

For noninteracting systems the Landauer-Büttiker (LB) formalism has
enjoyed great success as a tool for the description of transport processes
in the steady-state regime, when a long time has passed after the
switch on of a constant bias \citep{LANDAUER1957,LANDAUER1970,LandauerButtiker1982,BUTTIKER1990,BUTTIKER1992}.
This success may be attributed to its formal and conceptual simplicity,
and a set of equations for the steady-state current that can be combined
with DFT calculations to access the electronic properties of a system\citep{diventra2001DFT}.
It is useful for the description of transport processes in the ballistic
regime, where the time separating electron-electron interaction events
exceeds the time taken for an electron to cross the molecular region.
The LB formalism is convenient for the description of several features
of transport processes accessible to experimentalists: temperature-dependence\citep{diventra2001temp},
quantization of the conductance\citep{vanruitenbeek1995,vanruitenbeek2003},
exponential decay of conductance as a function of junction length\citep{vanruitenbeek2003,venkataraman2006},
and the shot and thermal noise associated with the current\citep{BUTTIKER1990,blantbutt}. 

In recent years, experimental work has increasingly focused on dynamical
properties of nanojunctions, in particular the AC current response
to a periodic voltage operating in the GHz or THz frequency range\citep{Burke2004,Li2004,Lin2009}.
It is therefore desirable to develop a formalism that will allow calculation
of the full time-dependent response of a molecular device to an arbitrary
time-dependent perturbation.

\textcolor{black}{Various approaches for going beyond the steady-state
exist. Neglecting interactions, these tend to involve extensions of
the LB formalism to time-dependent systems (TD-LB). In particular,
recent progress has been made by the construction of explicit scattering
state solutions for electrons incident on time-periodic non-stationary
scatterers\citep{MOSKALETS2012}. }

\textcolor{black}{The understanding of transport processes are fundamentally
problems in nonequilibrium statistical mechanics, because they typically
involve a sudden 'switch-on' of a coupling between subsystems or a
bias across subsystems which breaks the physical symmetry between
the system at $t=-\infty$ and $t=+\infty$. This means that if the
system was prepared in an equilibrium state in the distant past, it
will not relax back to its initial state in the far future. Historically,
there have been two main approaches to this kind of switch-on problem.
In }\textit{\textcolor{black}{partitioned}}\textcolor{black}{{} approaches\citep{CAROLI1,CAROLI2,JAUHO1993,JAUHO1994,diventra}
the system consists initially of decoupled leads with different chemical
potentials $\mu_{\alpha}$. At the initial time $t_{0}$, an idealized
coupling between C and the leads is added to the Hamiltonian. The
presence of the coupling enables electrons to hop between the $\alpha-C$
regions in the direction of decreasing chemical potential, }and the
LB formula is derived as the long-time limit of the resulting electron
current. This means that the appropriate electronic couplings are
not present in the equilibrium Hamiltonian prior to the switch-on,
and that embedding self-energies cannot be defined for the system
in equilibrium. However, this is not important for deriving the steady
state expression for the current as it is assumed that at long times
all preparation-dependent effects have been washed out. Working within
a partitioned approach, Jauho et. al. obtained a closed $\left(\omega,t\right)$-space
integral expression for the TD current response to an arbitrary time
varying bias\citep{JAUHO1993,JAUHO1994}. They assumed that the decoupled
leads were prepared at $t_{0}=-\infty$ with an appropriate time dependent
bias and then, immediately after, the coupling between the leads and
the central system was enabled. Naturally, this approach has a problem
as the bias and the coupling are being established at the same time
in the distant past. Therefore, the correctness of this approach rests
on an assumption that the system arrives at the correct non-equilibrium
state at times $t>0$ after a semi-infinite development from $-\infty$,
which cannot be guaranteed as the perturbation due to the coupling
\emph{and} the bias were added together. Also, in terms of applications,
the partitioned approach is limited to a periodic bias only; for instance,
the 'switch on' effect can only be modelled by considering periodic
bias pulses well separated in time. However, a finite time between
pulses prevents the system from equilibrating properly. However, in
a real experiment, one switches on the bias, not the coupling between
regions. The coupling will be fully present at equilibrium, so that
a single inverse temperature $\beta$ and single chemical potential
$\mu$ are defined for all regions of the junction. \textit{Partition-free}
approaches\citep{CINI1980,StefALm2004,RVL2014} recognise this fact,
and therefore they effectively diverge from partitioned approaches
in their choice of initial density matrix $\mathbf{\mathbf{\rho}_{0}}$
for the many-body dynamics. After $t_{0}$, the same Hamiltonian is
used to propagate the system in either of the two approaches, so if
all information on the preparation of the system is lost in the long-time
limit, one would expect the same steady state to be reached regardless
of the choice of the initial density matrix. 

However, whereas the equilibrium initial distribution in the partition-free
approach is well-defined and leads to a physical transient current,
any transients in the partitioned approach cannot be interpreted physically,
as they result from the junction's 'memory' of its fictitious initial
state. Important progress was made in the direction of a fully time-dependent
partition-free approach by Stefanucci and Almbladh, who proved theorems
of asymptotic equivalence between the two approaches and obtained
an integral formula for the linear current response to a TD bias within
the Wide-Band-Limit Approximation (WBLA)\citep{StefALm2004}. Recently,
Tuovinen \textit{\textcolor{black}{et. al.}} obtained a closed frequency
integral for the exact time-dependent current response to a static
bias, within a partition-free approach, making it increasingly possible
to theoretically investigate short-time effects in nanoscale systems\citep{RVLS2013,RVL2013}.
Their formula clearly exhibits decaying modes in addition to a steady-state
part, and has been applied to the study of transport through ring-shaped
molecular junctions\citep{RVL2013}, and more recently, graphene nanoribbons\citep{RVL2014}.

Here we develop an analytic approach whereby a TD-LB formula for a
current response to an\emph{ arbitrary }time-dependent bias in a multi-lead
system, including the 'switch-on' effect, is derived, for any tight-binding
Hamiltonian. The final formula involves only slightly more computational
cost than the corresponding LB formula for the steady-state current.
In our formulation we extend the partition-free approach of \citep{RVL2013,RVLS2013},
in that no assumptions are made concerning the particular time dependence
of the bias applied to each of the leads. The analytical result is
obtained within the WBLA by integrating exactly the Kadanoff-Baym
equations for the Green's function of the central region. This work
incorporates the formulae of Jauho et. al. and Tuovinen et. al. as
special cases. Specifically, the former emerges as the long time limit
of a more general expression that contains the transients. The latter
is the constant bias limit of our more general expression containing
time integrals of the bias. In this way we show that an analytical
expression for the fully non-linear current response to an arbitrary
TD bias at all times subsequent to $t_{0}$ is possible within the
WBLA.

Note in passing that a numerical approach to the solution of the embedded
Kadanoff-Baym equations for systems that include an explicit electron-electron
interaction term in their Hamiltonian within the Second Born (2B)
and GW approximations has also been recently developed\citep{RVLS2006,RVL2006,RVL2009,RVL2014}.
Extensions of the tight-binding approach to a DFT-Green's function
formalism in the time domain have also been performed. In these approaches,
the system is prepared in an equilibrium state whose electronic structure
is obtained from \textit{ab initio} calculations, before the time-dependent
perturbation is applied\citep{Yang2010}. We do not take these approaches
here as they can only be facilitated numerically. We believe that
a\textcolor{green}{{} }\textcolor{black}{great deal of physics is readily
accessible with a simple tight-binding model and a simplified treatement
of the coupling (the essence of the WBLA). Some physical effects that
one would expect to observe, for example in the case of a periodic
bias, include asymmetry of the current signal about the voltage peak,
and a quickly oscillating transient, the so-called 'ringing' of the
current immediately after the switch-on. In the NEGF-based formalism
for calculating the current response to an arbitrary time-dependent
bias these effects are clearly observable, amongst others. We emphasize
that our formalism is not limited to the case of a periodic bias,
and therefore offers an alternative to the non-stationary S-matrix
approaches which rely on the Floquet theorem. }

The paper is organised as follows: section 2 contains a description
of the model Hamiltonian, the derivation of self-energies and all
components of the Green's functions required for the current. In particular,
a novel formula for the exact lesser Green's function of two times
is presented here. From this, an $\left(\omega,t\right)$ integral
is derived for the current through one of the leads, and we proceed
to show how it reduces to known results in the long-time and static
bias limits. Note that some steps of our calculation are similar to
those done in \citep{RVL2013}. In these cases we do not give much
detail; instead we briefly outline a few essential steps and state
the final results. In Section \ref{sec:Results} we present the results
of current calculations on a quantum dot central region in response
to a periodic external bias. Appendices A to C contain details of
the calculation of the lesser Green's function and of the current.

\section{Theory \label{sec:Theory}}

\subsection{Hamiltonian on the contour}

The Kadanoff-Baym or Nonequilibrium Green's function (NEGF) approach
to electron transport is a tool for the description of the time evolution
ensemble averages of quantum systems; in particular, it can conveniently
be used for calculating time dependent ensemble averages of particle
number densities in subsystems open to an environment. 

The most essential feature of this formalism is that of time evolution
on a contour defined in the complex time plane. Here, $\gamma$ refers
to the Konstantinov-Perel' contour, which is just the union of the
three sets of points: (i) $C_{-}$, the 'upper' contour, containing
points between $t_{0}+i0$ and $t+i0$; (ii) the lower' contour $C_{+}$
containing points along the line between $t-i0$ and $t_{0}-i0$,
and (iii) $C_{M}$ which is any contour in the complex plane between
two points $z_{a}$, $z_{b}$, such that $z_{b}-z_{a}=-i\beta$, where
$\beta=1/k_{B}T$ is the inverse temperature. Usually one chooses
$z_{a}=t_{0}-i0$ and $z_{b}=t_{0}-i\beta$, and connects them with
a vertical path on the contour as shown in Fig. \ref{fig:The-Konstantinov-Perel'-Contour}.
At 'times' that lie on $C_{M}$ the system exists in thermodynamic
equilibrium since this part of the contour originates from the initial
density operator $\widehat{\rho}_{0}=Z^{-1}e^{-\beta\left(\widehat{H}_{0}-\mu\widehat{N}\right)}$,
with $\mu$ being the chemical potential of the whole system and $\widehat{N}$
the particle number operator. The times on the contour are ordered
from $t_{0}+i0$ to $t_{0}-i\beta$ and then to $t_{0}-i\beta$ as
shown by the arrows in \ref{fig:The-Konstantinov-Perel'-Contour}.
In this paper we use the units in which $\hbar=1$. 

The variable $z$ will be used to indicate the 'time' along the contour,
and the system Hamiltonian $\hat{H}\left(z\right)$ must be specified
for all $z$. When used between z-variables, the symbols `$>/<$'
mean `later/earlier on the contour', where $z_{1}$ is later than
$z_{2}$ on the contour if a directed line drawn from the point $t_{0-}$
to $t_{0}-i\beta$ passes through $z_{2}$ before passing through
$z_{1}$. In general, the definition of the Hamiltonian will be different
on different parts of the contour, containing both equilibrium and
nonequilibrium parts; for instance, the non-equilibrium part contains
the bias whicb is missing in $\widehat{H}_{0}$ defined on the vertical
track. For the typical set-up of a quantum transport problem, we therefore
have the following Hamiltonian:

\begin{figure}
\begin{centering}
\includegraphics[height=5cm]{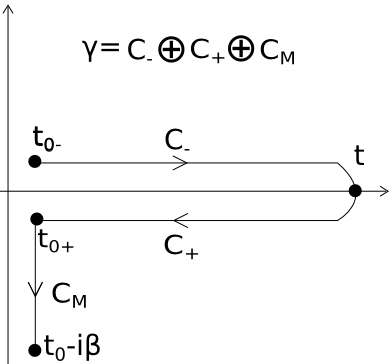}
\par\end{centering}

\caption{The Konstantinov-Perel' Contour $\gamma\equiv C_{-}\oplus C_{+}\oplus C_{M}$.
\label{fig:The-Konstantinov-Perel'-Contour}}
\end{figure}

\begin{equation}
\hat{H}\left(z\right)=\underset{k\alpha\sigma}{\sum}\varepsilon_{k\alpha}\left(z\right)\hat{d}_{k\alpha\sigma}^{\dagger}\hat{d}_{k\alpha\sigma}+\underset{mn\sigma}{\sum}T_{mn}\left(z\right)\hat{d}_{m\sigma}^{\dagger}\hat{d}_{n\sigma}+\underset{m,k\alpha\sigma}{\sum}\left[T_{m,k\alpha}\left(z\right)\hat{d}_{m\sigma}^{\dagger}\hat{d}_{k\alpha\sigma}+T_{k\alpha,m}\left(z\right)\hat{d}_{k\alpha\sigma}^{\dagger}\hat{d}_{m\sigma}\right]\label{eq:Hamiltonian}
\end{equation}
The first term in this expression corresponds to the sum of the Hamiltonians
of the reservoirs/leads, where $\alpha$ labels the lead, and $k$
labels the $k$-th eigenstate of this lead. The second term corresponds
to the Hamiltonian of the central region $C$ and hence refers to
electron hopping events within this region with indices $n$ and $m$
labeling eigenstates there. Finally, the third term describes the
coupling of the leads and the central system with the corresponding
matrix elements $T_{m,k\alpha}$, and $\sigma$ denotes the spin degree
of freedom of the electrons. Correspondingly, $\hat{d}_{k\alpha\sigma}$,
$\hat{d}_{m\sigma}$ and $\hat{d}_{k\alpha\sigma}^{\dagger}$, $\hat{d}_{m\sigma}^{\dagger}$
are destruction and creation operators of the leads and the central
system.

In the following, bold letters will be used for matrices. We shall
omit the spin index for simplicity of notation; it can be trivially
introduced back in the final results if desired. It is useful at this
point to introduce a matrix $\mathbf{h}(z)$, defined on the basis
of the whole multi-terminal system, so that the Hamiltonian (\ref{eq:Hamiltonian})
can be rewritten simply as 
\begin{equation}
\widehat{H}(z)=\sum_{ij}h_{ij}\hat{d}_{i}^{\dagger}\hat{d}_{j}\label{eq:Hamiltonian-simple}
\end{equation}
where we sum over all orbitals $i,j$ of the leads and the central
system, $h_{ij}$ are elements of the matrix $\mathbf{h}$, and $d_{i}$,
$d_{i}^{\dagger}$ correspond to the operators of either the leads
(if $i\equiv k\alpha$) or $C$ (if $i\equiv n$). It is also convenient
to introduce blocks of the matrix $\mathbf{h}$, projected onto the
molecule or reservoir subspaces, i.e. $\mathbf{h}_{\alpha\alpha^{\prime}}(z)$
is the block of $\mathbf{h}(z)$ with matrix elements $\left[\mathbf{h}_{\alpha\alpha'}\right]_{kk'}=\delta_{\alpha\alpha'}\delta_{kk'}\varepsilon_{k\alpha}(z)$
(note that there is no interaction between leads, so that $\mathbf{h}_{\alpha\alpha^{\prime}}=\delta_{\alpha\alpha^{\prime}}\mathbf{h}_{\alpha\alpha}$),
$\mathbf{h}_{\alpha C}$ is the block of $\mathbf{h}(z)$ with matrix
elements $\left[\mathbf{h}_{\alpha C}\right]_{km}=T_{k\alpha,m}(z)$,
and $\mathbf{h}_{CC}$ is the block of $\mathbf{h}(z)$ with matrix
elements $\left[\mathbf{h}_{CC}\right]_{mn}=T_{mn}(z)$. We assume
that prior to the switch on at time $t_{0}$ the whole electron system
was in thermodynamic equilibrium characterised by the unique chemical
potential $\mu$ and inverse temperature $\beta$. Then at $t_{0}$
each lead $\alpha$ is subjected to the arbitrary bias potential $V_{\alpha}(t)$.
Then on different parts of the contour $\gamma$ the elements of the
matrix $\mathbf{h}(z)$ are defined as follows:

\begin{equation}
\left[\mathbf{h}_{\alpha\alpha}\left(z\right)\right]_{kk'}=\left\{ \begin{array}{c}
\left(\varepsilon_{k\alpha}+V_{\alpha}\left(t\right)\right)\delta_{kk'},\; z\equiv t\in C_{-}\oplus C_{+}\\
\left(\varepsilon_{k\alpha}-\mu\right)\delta_{kk'},\; z\in C_{M}
\end{array}\right.\label{eq:H_for_leads}
\end{equation}
\begin{equation}
\left[\mathbf{h}_{C\alpha}\left(z\right)\right]_{mk}=T_{mk,\alpha},\; z\in\gamma\label{eq:H_for_lead-C}
\end{equation}
\begin{equation}
\left[\mathbf{h}_{CC}\left(z\right)\right]_{mn}=\left\{ \begin{array}{c}
T_{mn},\; z\in C_{-}\oplus C_{+}\\
T_{mn}-\mu\delta_{mn},\; z\in C_{M}
\end{array}\right.\label{eq:H_for_central}
\end{equation}

This Hamiltonian corresponds to the switch-on of a spatially uniform
bias in each reservoir. 

\textcolor{black}{One note is in order here. The experimentalist in
such systems typically has control over the voltage passed across
the nanojunction. At some initial time $t_{0}$, the switch-on of
a bias means that energy levels in lead $\alpha$ are shifted by some
external bias, $V_{\alpha}\left(t\right)$. In general, an external
field will cause a rearrangement of electrons in the junction and
leads. In particular, the electrons in the conducting leads will screen
the external field so that the electric field inside the conductor
equals zero and its electric potential is uniform throughout. This
leads to confinement of the drop in potential to the region C, and
means that $V_{\alpha}\left(t\right)$ should not be interpreted as
$eV\left(t\right)$, where $V\left(t\right)$ is the actual voltage
across the whole macroscopic lead. Rather, it should be equal to the
sum of $eV\left(t\right)$ and the screening potential induced by
the switch-on. However, the central region $C$ can always be chosen
sufficiently large that a part of each lead neighbouring $C$, which
is mostly affected by the charge redistribution is incorporated into
$C$. In this case we may assume that the potential $V_{\alpha}\left(t\right)$
is uniform across the lead $\alpha$ and is equal to the bias voltage
applied to this particular lead.}

\subsection{Greens function components}

The one-particle Green's function on the contour $\gamma$ is defined
in the usual way as: 

\[
G_{ij}\left(z_{1},z_{2}\right)\equiv-i\frac{\mbox{Tr}\left\{ e^{-\beta\widehat{H}_{M}}\hat{T}_{\gamma}\left[\hat{d}_{H,i}\left(z_{1}\right)\hat{d}_{H,j}^{\dagger}\left(z_{2}\right)\right]\right\} }{\mbox{Tr}\left(e^{-\beta\widehat{H}_{M}}\right)}
\]
where $\widehat{T}_{\gamma}$ is the 'time' ordering operator on the
contour $\gamma$, $\widehat{H}_{M}$ is the Hamiltonian on its vertical
part, and the subscript $H$ by the creation and annihilation operators
means that these are considered in the Heisenberg picture on the contour.
The elements $G_{ij}$ of the Green's function form a matrix $\mathbf{G}$
defined on the whole space of orbitals of all leads and the central
region; correspondingly, one can introduce diagonal, $\mathbf{G}_{CC}$
and $\mathbf{G}_{\alpha\alpha}$, as well as non-diagonal, $\mathbf{G}_{C\alpha}$
and $\mathbf{G}_{\alpha C}$, blocks of this matrix. The Green's function
$\mathbf{G}_{CC}$ for the central region satisfies the equations
of motion:
\begin{equation}
\left[i\frac{d}{dz_{1}}-\mathbf{h}_{CC}\left(z_{1}\right)\right]\mathbf{G}_{CC}\left(z_{1},z_{2}\right)=\mathbf{1}_{CC}\delta\left(z_{1},z_{2}\right)+\int_{\gamma}d\bar{z}\,\mathbf{\Sigma}_{CC}\left(z_{1},\bar{z}\right)\mathbf{G}_{CC}\left(\bar{z},z_{2}\right)\label{eq:EoM-for-G_CC-1}
\end{equation}
\begin{equation}
\mathbf{G}_{CC}\left(z_{1},z_{2}\right)\left[-i\frac{\overleftarrow{d}}{dz_{2}}-\mathbf{h}_{CC}\left(z_{2}\right)\right]=\mathbf{1}_{CC}\delta\left(z_{1},z_{2}\right)+\int_{\gamma}d\bar{z}\,\mathbf{G}_{CC}\left(z_{1},\bar{z}\right)\mathbf{\Sigma}_{CC}\left(\bar{z},z_{2}\right)\label{eq:EoM-for-G_CC-2}
\end{equation}
where $\mathbf{1}_{CC}$ is the unit matrix defined on the orbitals
of $C$, and 
\begin{equation}
\mathbf{\Sigma}_{CC}\left(z_{1},z_{2}\right)=\underset{\alpha}{\sum}\mathbf{h}_{C\alpha}\left(z_{1}\right)\mathbf{g}_{\alpha\alpha}\left(z_{1},z_{2}\right)\mathbf{h}_{\alpha C}\left(z_{2}\right)\label{eq:self-energy-Def}
\end{equation}
is the matrix of the embedded self-energy, while the non-diagonal
matrix blocks of the Green's function are given by the integrals on
the contour:
\begin{equation}
\mathbf{G}_{\alpha C}\left(z_{1},z_{2}\right)=\int_{\gamma}d\bar{z}\,\mathbf{g}_{\alpha\alpha}\left(z_{1},\bar{z}\right)\mathbf{h}_{\alpha C}\left(\bar{z}\right)\mathbf{G}_{CC}\left(\bar{z},z_{2}\right)\label{eq:off-diag-GF}
\end{equation}
\begin{equation}
\mathbf{G}_{C\alpha}\left(z_{1},z_{2}\right)=\int_{\gamma}d\bar{z}\,\mathbf{G}_{CC}\left(z_{1},\bar{z}\right)\mathbf{h}_{C\alpha}\left(\bar{z}\right)\mathbf{g}_{\alpha\alpha}\left(\bar{z},z_{2}\right)\label{eq:off-diag-GF-1}
\end{equation}
Here $\mathbf{g}_{\alpha\alpha}\left(z_{1},z_{2}\right)$ is the Green's
function of an isolated lead $\alpha$ corresponding to the Hamiltonian
with the matrix block $\mathbf{h}_{\alpha\alpha}(z)$; note that the
latter contains the term with the bias.

Depending on the positions of the 'time' arguments $z_{1}$ and $z_{2}$
on $\gamma$, special notations are normally used for the Green's
functions defining its various components: if $z_{1},z_{2}$ belong
to either of the horizontal parts of $\gamma$ with $z_{1}>z_{2}$
or $z_{1}<z_{2}$, then the greater, $\mathbf{G^{>}}\left(t_{1},t_{2}\right)$,
or lesser, $\mathbf{G}^{<}\left(t_{1},t_{2}\right)$, components are
defined, respectively; if one of the times lies on the vertical part
and another on a horizontal one, then the right, $\mathbf{G}^{\urcorner}\left(t_{1},\tau_{2}\right)$,
and left, $\mathbf{G}^{\ulcorner}\left(\tau_{1},t_{2}\right)$, components
are introduced, respectively, where $t_{1}$ or $t_{2}$ lie on the
horizontal parts while another arguments $\tau_{1}$ or $\tau_{2}$
correspond to the 'times' $t_{0}-i\tau_{1}$ or $t_{0}-i\tau_{2}$
on the vertical part of $\gamma$; finally, it is also convenient
to consider the case when both times lie on the vertical part of $\gamma$
in which case $\mathbf{G}\left(z_{1},z_{2}\right)$ coincides with
the Matsubara Green's function $\mathbf{G}^{M}\left(\tau_{1},\tau_{2}\right)$.
In addition to the objects defined above, on the horizontal part of
$\gamma$ we define, respectively, the retarded and advanced Green's
functions:

\begin{eqnarray*}
\mathbf{G}^{r}\left(t_{1},t_{2}\right) & = & \theta\left(t_{1}-t_{2}\right)\left[\mathbf{G}^{>}\left(t_{1},t_{2}\right)-\mathbf{G}^{<}\left(t_{1},t_{2}\right)\right]\\
\mathbf{G}^{a}\left(t_{1},t_{2}\right) & = & -\theta\left(t_{2}-t_{1}\right)\left[\mathbf{G}^{>}\left(t_{1},t_{2}\right)-\mathbf{G}^{<}\left(t_{1},t_{2}\right)\right]
\end{eqnarray*}
where $\theta\left(t\right)$ is the unit step function.

\subsection{Self-energies in the WBLA}

Depending on where the two 'time' arguments in the self-energy $\mathbf{\Sigma}_{CC}\left(z_{1},z_{2}\right)$
lie, various components of it can also be defined. These require calculation
of the corresponding components of the Green's function $\mathbf{g}_{\alpha\alpha}\left(z_{1},z_{2}\right)$
of the isolated lead $\alpha$. The time-dependent part, $\sum_{k}V_{\alpha}(t)\hat{N}_{k\alpha}$
(where $\hat{N}_{k\alpha}=\hat{d}_{k\alpha}^{\dagger}\hat{d}_{k\alpha}$
is the number operator), in the Hamiltonian of the lead $\alpha$
commutes with the rest of its Hamiltonian and hence the time ordering
can be omitted when calculating the corresponding evolution operator.
Hence, the expressions for the creating and annihilation operators
in the Heisenberg representation follow immediately: if $z\equiv t\in C_{\pm}$,
then

\[
\hat{d}_{k\alpha}\left(t\right)=\hat{d}_{k\alpha}e^{-i\phi_{k\alpha}\left(t,t_{0}\right)}=\left[\hat{d}_{k\alpha}^{\dagger}\left(t\right)\right]^{\dagger}
\]
where it is convenient to introduce a function
\[
\phi_{k\alpha}\left(t,t_{0}\right)=\left(t-t_{0}\right)\epsilon_{k\alpha}+\intop_{t_{0}}^{t}V_{\alpha}\left(\tau\right)d\tau=\left(t-t_{0}\right)\epsilon_{k\alpha}+\psi_{\alpha}\left(t,t_{0}\right)\;;
\]
if $z=t_{0}-i\tau\in C_{M}$, then the creation/annihilation operators
are no longer Hermitian conjugates of each other:

\[
\hat{d}_{k\alpha}\left(t_{0}-i\tau\right)=\hat{d}_{k\alpha}e^{-\left(\varepsilon_{k\alpha}-\mu\right)\tau}
\]
\[
\hat{d}_{k\alpha}^{\dagger}\left(t_{0}-i\tau\right)=\hat{d}_{k\alpha}^{\dagger}e^{\left(\varepsilon_{k\alpha}-\mu\right)\tau}
\]

Using these operators and the definition of the Green's functions,
different components of the lead Green's functions are obtained:
\[
\left[\mathbf{g}_{\alpha\alpha}^{r}\left(t_{1},t_{2}\right)\right]_{kk'}=-i\theta\left(t_{1}-t_{2}\right)\delta_{kk'}e^{-i\phi_{k\alpha}\left(t_{1},t_{2}\right)}
\]
\[
\left[\mathbf{g}_{\alpha\alpha}^{a}\left(t_{1},t_{2}\right)\right]_{kk'}=i\theta\left(t_{2}-t_{1}\right)\delta_{kk'}e^{-i\phi_{k\alpha}\left(t_{1},t_{2}\right)}
\]
\[
\left[\mathbf{g}_{\alpha\alpha}^{<}\left(t_{1},t_{2}\right)\right]_{kk'}=i\delta_{kk'}f\left(\varepsilon_{k\alpha}-\mu\right)e^{-i\phi_{k\alpha}\left(t_{1},t_{2}\right)}
\]
\[
\left[\mathbf{g}_{\alpha\alpha}^{>}\left(t_{1},t_{2}\right)\right]_{kk'}=-i\delta_{kk'}\left[1-f\left(\varepsilon_{k\alpha}-\mu\right)\right]e^{-i\phi_{k\alpha}\left(t_{1},t_{2}\right)}
\]
Here $f(\omega)=\left(e^{\beta\omega}+1\right)^{-1}$ is the Fermi
function. In the case of the Matsubara component, it is convenient
to expand it into a Fourier series, which is possible due to the boundary
conditions, $\mathbf{g}_{\alpha\alpha}^{M}(\tau,0)=-\mathbf{g}_{\alpha\alpha}^{M}(\tau,\beta)$
and $\mathbf{g}_{\alpha\alpha}^{M}(0,\tau)=-\mathbf{g}_{\alpha\alpha}^{M}(\beta,\tau)$:
\[
\left[\mathbf{g}_{\alpha\alpha}^{M}\left(\tau_{1},\tau_{2}\right)\right]_{kk'}=\delta_{kk'}\frac{i}{\beta}\underset{q}{\sum}\frac{e^{-\omega_{q}\left(\tau_{1}-\tau_{2}\right)}}{\omega_{q}-\varepsilon_{k\alpha}+\mu}
\]
where $\omega_{q}=i\pi\left(2q+1\right)/\beta$ are Matsubara frequencies
and the $q$ summation is run over all negative and positive integers.
Then, the right and left components can also be written via the Matsubara
sums: 
\[
\left[\mathbf{g}_{\alpha\alpha}^{\urcorner}\left(t,\tau\right)\right]_{kk'}=\left[\mathbf{g}_{\alpha\alpha}^{M}\left(0,\tau\right)\right]_{kk'}e^{-i\phi_{k\alpha}\left(t,t_{0}\right)}=\delta_{kk'}\frac{i}{\beta}e^{-i\phi_{k\alpha}\left(t,t_{0}\right)}\underset{q}{\sum}\frac{e^{\omega_{q}\tau}}{\omega_{q}-\varepsilon_{k\alpha}+\mu}
\]
\[
\left[\mathbf{g}_{\alpha\alpha}^{\ulcorner}\left(\tau,t\right)\right]_{kk'}=\left[\mathbf{g}_{\alpha\alpha}^{M}\left(\tau,0\right)\right]_{kk'}e^{i\phi_{k\alpha}\left(t,t_{0}\right)}=\delta_{kk'}\frac{i}{\beta}e^{i\phi_{k\alpha}\left(t,t_{0}\right)}\underset{q}{\sum}\frac{e^{-\omega_{q}\tau}}{\omega_{q}-\varepsilon_{k\alpha}+\mu}
\]

All the components of the self-energy can now be obtained from Eq.
(\ref{eq:self-energy-Def}). To obtain the retarded component, we
shall Fourier transform that part of the expression since it depends
only on the time-difference $t_{1}-t_{2}$: 
\[
\left[\mathbf{\Sigma}_{CC}^{r}\left(t_{1},t_{2}\right)\right]_{mn}=\sum_{\alpha}e^{-i\psi_{\alpha}\left(t_{1},t_{2}\right)}\int\frac{d\omega}{2\pi}e^{i\omega\left(t_{1}-t_{2}\right)}\underset{k}{\sum}\frac{T_{m,k\alpha}T_{k\alpha,n}}{\omega+i0-\varepsilon_{k\alpha}}
\]
\[
=\sum_{\alpha}e^{-i\psi_{\alpha}\left(t_{1},t_{2}\right)}\int\frac{d\omega}{2\pi}e^{i\omega\left(t_{1}-t_{2}\right)}\left[\Lambda_{\alpha,mn}\left(\omega\right)-\frac{i}{2}\Gamma_{\alpha,mn}\left(\omega\right)\right]
\]
where (the symbol $\mathcal{P}$ corresponds to the Cauchy principal
part): 
\[
\Lambda_{\alpha,mn}\left(\omega\right)=\mathcal{P}\int\frac{d\omega^{\prime}}{2\pi}\frac{\Gamma_{\alpha,nm}\left(\omega^{\prime}\right)}{\omega-\omega^{\prime}}
\]
\[
\Gamma_{\alpha,mn}\left(\omega\right)=2\pi\sum_{k}T_{m,k\alpha}T_{k\alpha,n}\delta\left(\omega-\epsilon_{k\alpha}\right)
\]
In the WBLA it is assumed that the energy range of the transmission
channel in the leads is so wide that the coupling between every lead
state and a given molecular orbital is independent of the energy of
the lead state; hence, the $\omega$ dependence of the level width
matrix $\mathbf{\Gamma}_{\alpha}=\left\Vert \Gamma_{\alpha,mn}\left(\omega\right)\right\Vert $
is neglected, $\Gamma_{\alpha,mn}\left(\omega\right)\simeq\Gamma_{\alpha,mn}$,
in which case the level shift matrix becomes exactly zero, $\mathbf{\Lambda}_{\alpha}\left(\omega\right)=0$.
This is a good approximation in systems for which transport takes
place at energies close to the Fermi level, and in cases where the
matrices $\mathbf{\Lambda}_{\alpha}\left(\omega\right)$ and $\mathbf{\Gamma}_{\alpha}\left(\omega\right)$
vary slowly with $\omega$. This approximation finally gives a simple
analytical result: 
\begin{equation}
\left[\mathbf{\Sigma}_{CC}^{r}\left(t_{1},t_{2}\right)\right]_{mn}=-\frac{i}{2}\delta\left(t_{1}-t_{2}\right)\Gamma_{mn}\label{eq:self-energy-R}
\end{equation}
where $\Gamma_{mn}=\sum_{\alpha}\Gamma_{\alpha,mn}$ is the total
level width of the junction. Similarly, 
\begin{equation}
\left[\mathbf{\Sigma}_{CC}^{a}\left(t_{1},t_{2}\right)\right]_{mn}=\frac{i}{2}\delta\left(t_{1}-t_{2}\right)\Gamma_{mn}\label{eq:self-energy-A}
\end{equation}

In the WBLA, the lesser, greater, right and left self-energy components
are obtained in a similar manner. We shall give them here for completeness:
\begin{equation}
\left[\mathbf{\Sigma}_{CC}^{<}\left(t_{1},t_{2}\right)\right]_{mn}=i\left[\sum_{\alpha}\Gamma_{\alpha,mn}e^{-i\psi_{\alpha}\left(t_{1},t_{2}\right)}\right]\int\frac{d\omega}{2\pi}f\left(\omega-\mu\right)e^{-i\omega\left(t_{1}-t_{2}\right)}\label{eq:Greater-self-energy}
\end{equation}
\begin{equation}
\left[\mathbf{\Sigma}_{CC}^{>}\left(t_{1},t_{2}\right)\right]_{mn}=-i\left[\sum_{\alpha}\Gamma_{\alpha,mn}e^{-i\psi_{\alpha}\left(t_{1},t_{2}\right)}\right]\int\frac{d\omega}{2\pi}\left[1-f\left(\omega-\mu\right)\right]e^{-i\omega\left(t_{1}-t_{2}\right)}\label{eq:Lesser-self-energy}
\end{equation}
\begin{equation}
\left[\mathbf{\Sigma}_{CC}^{\ulcorner}\left(\tau,t\right)\right]_{mn}=\frac{i}{\beta}\left[\sum_{\alpha}\Gamma_{\alpha,mn}e^{i\psi_{\alpha}\left(t,t_{0}\right)}\right]\underset{q}{\sum}e^{-\omega_{q}\tau}\int\frac{d\omega}{2\pi}\frac{e^{i\omega\left(t-t_{0}\right)}}{\omega_{q}-\omega+\mu}\label{eq:left-self-energy}
\end{equation}
\begin{equation}
\left[\mathbf{\Sigma}_{CC}^{\urcorner}\left(t,\tau\right)\right]_{mn}=\frac{i}{\beta}\left[\sum_{\alpha}\Gamma_{\alpha,mn}e^{-i\psi_{\alpha}\left(t,t_{0}\right)}\right]\underset{q}{\sum}e^{\omega_{q}\tau}\int\frac{d\omega}{2\pi}\frac{e^{-i\omega\left(t-t_{0}\right)}}{\omega_{q}-\omega+\mu}\label{eq:right-self-energy}
\end{equation}

The Matsubara self-energy is obtained similarly with the help of the
identity 
\[
\int\frac{d\omega}{2\pi}\frac{1}{\omega_{q}-\omega+\mu}=\frac{i}{2}\xi_{q}
\]
Here $\xi_{q}=+1$ when $Im\left(\omega_{q}\right)>0$ and $\xi_{q}=-1$
when $Im\left(\omega_{q}\right)<0$. This finally gives:$ $
\begin{equation}
\left[\mathbf{\Sigma}_{CC}^{M}\left(\tau_{1},\tau_{2}\right)\right]_{mn}=-\frac{\Gamma_{mn}}{2\beta}\underset{q}{\sum}\xi_{q}e^{-\omega_{q}\left(\tau_{1}-\tau_{2}\right)}\label{eq: Matsubara-self-energy}
\end{equation}
One can see that expressions for the lesser, greater, right and left
componenets of the self-energy differ from those obtained in \citep{RVL2013,RVLS2013}
for the case of the switched on constant bias; the expressions for
the Matsubara, advanced and retarded components remain the same.

\subsection{Integration of the Kadanoff-Baym equations}

\subsubsection{Matsubara Green's function}

Projecting Eq. (\ref{eq:EoM-for-G_CC-1}) on the vertical track of
the contour $\gamma$ and applying the Langreth rules\citep{Langreth1972,Langreth1976},
the corresponding equation of motion for the Matsubara Green's function
$\mathbf{G}_{CC}^{M}\left(\tau_{1},\tau_{2}\right)$ is obtained:
\begin{equation}
\left[-\frac{d}{d\tau_{1}}-\mathbf{h}_{CC}^{M}\right]\mathbf{G}_{CC}^{M}\left(\tau_{1},\tau_{2}\right)=i\mathbf{1}_{CC}\delta\left(\tau_{1}-\tau_{2}\right)+\left(\mathbf{\Sigma}_{CC}^{M}\star\mathbf{G}_{CC}^{M}\right)_{\left(\tau_{1},\tau_{2}\right)}\label{eq:Matsubara-EoM}
\end{equation}
where $\mathbf{h}_{CC}^{M}=\mathbf{h}_{CC}-\mu\mathbf{1}_{CC}$ is
the Hamiltonian matrix of the central region on the vertical track
of the contour $\gamma$ with $\left(\mathbf{h}_{CC}\right)_{mn}=T_{mn}$,
while the star in the last term corresponds to the imaginary time
convolution integral \citep{RVLS2013}:
\begin{equation}
\left(\mathbf{\Sigma}_{CC}^{M}\star\mathbf{G}_{CC}^{M}\right)_{\left(\tau_{1},\tau_{2}\right)}\equiv-i\int_{0}^{\beta}d\tau\mathbf{\Sigma}_{CC}^{M}\left(\tau_{1},\tau\right)\mathbf{G}_{CC}^{M}\left(\tau,\tau_{2}\right)\label{eq:Def-of-star-product}
\end{equation}
Expanding the self-energy $\mathbf{\Sigma}_{CC}^{M}\left(\tau,\tau^{\prime}\right)$,
the Green's function $\mathbf{G}_{CC}^{M}\left(\tau,\tau^{\prime}\right)$
and the delta function $\delta\left(\tau_{1}-\tau_{2}\right)$ into
the Matsubara sums, one immediately obtains:

\begin{equation}
\mathbf{G}_{CC}^{M}\left(\tau_{1},\tau_{2}\right)=\frac{i}{\beta}\sum_{q}e^{-\omega_{q}\left(\tau_{1}-\tau_{2}\right)}\left\{ \begin{array}{c}
\left[\left(\omega_{q}+\mu\right)\mathbf{1}_{CC}-\mathbf{h}_{CC}^{eff}\right]^{-1},\; Im\left(\omega_{q}\right)>0\\
\left[\left(\omega_{q}+\mu\right)\mathbf{1}_{CC}-\left(\mathbf{h}_{CC}^{eff}\right)^{\dagger}\right]^{-1},\; Im\left(\omega_{q}\right)<0
\end{array}\right.\label{eq:Matsubara-GF-expansion}
\end{equation}
\begin{equation}
\mathbf{G}_{CC}^{M}\left(\omega_{q}\right)=\left\{ \begin{array}{c}
\left[\left(\omega_{q}+\mu\right)\mathbf{1}_{CC}-\mathbf{h}_{CC}^{eff}\right]^{-1},\; Im\left(\omega_{q}\right)>0\\
\left[\left(\omega_{q}+\mu\right)\mathbf{1}_{CC}-\left(\mathbf{h}_{CC}^{eff}\right)^{\dagger}\right]^{-1},\; Im\left(\omega_{q}\right)<0
\end{array}\right.\label{eq:Matsubara-GF-components}
\end{equation}
where $\mathbf{h}_{CC}^{eff}=\mathbf{h}_{CC}-i\mathbf{\Gamma}_{CC}/2$
is the ``effective'' Hamiltonian of region $C$. As expected, the
Matsubara Green's function is the same as in \citep{RVL2013}. This
is to be expected as it corresponds to the preparation of the system
prior to the switch on of the bias.

\subsubsection{Right and left Green's functions}

Projecting Eq. (\ref{eq:EoM-for-G_CC-1}) on the right component $ $(i.e.
$t_{1}\in C_{-}\oplus C_{+}$, $t_{2}=t_{0}-i\tau_{2}\in C_{M}$)
and using the Langreth rules again, the following equation of motion
for the right Green's function is obtained:
\[
\left(i\frac{d}{dt_{1}}-\mathbf{h}_{CC}\right)\mathbf{G}_{CC}^{\urcorner}\left(t_{1},\tau_{2}\right)=\left(\mathbf{\Sigma}_{CC}^{r}\cdot\mathbf{G}_{CC}^{\urcorner}+\mathbf{\Sigma}_{CC}^{\urcorner}\star\mathbf{G}_{CC}^{M}\right)_{\left(t_{1},\tau_{2}\right)}
\]
where, following notations introduced in \citep{RVLS2013}, we defined
the dot-convolution (real time axis) integral 
\begin{equation}
\left(\mathbf{\Sigma}_{CC}^{r}\cdot\mathbf{G}_{CC}^{\urcorner}\right)_{\left(t,\tau\right)}=\int_{t_{0}}^{\infty}dt_{1}\mathbf{\Sigma}_{CC}^{r}\left(t,t_{1}\right)\mathbf{G}_{CC}^{\urcorner}\left(t_{1},\tau\right)\label{eq:Def-of-dot-product}
\end{equation}
Using expression for the retarded self-energy (\ref{eq:self-energy-R}),
the equation of motion is manipulated into:
\[
\left(i\frac{d}{dt_{1}}-\mathbf{h}_{CC}^{eff}\right)\mathbf{G}_{CC}^{\urcorner}\left(t_{1},\tau_{2}\right)=\left(\mathbf{\Sigma}_{CC}^{\urcorner}\star\mathbf{G}_{CC}^{M}\right)_{\left(t_{1},\tau_{2}\right)}
\]
which can be solved for the right Green's function taking into account
the appropriate boundary conditions $\mathbf{G}_{CC}^{\urcorner}\left(t_{0},\tau\right)=\mathbf{G}_{CC}^{M}\left(0^{-},\tau\right)$:
\begin{equation}
\mathbf{G}_{CC}^{\urcorner}\left(t,\tau\right)=e^{-i\mathbf{h}_{CC}^{eff}\left(t-t_{0}\right)}\left[\mathbf{G}_{CC}^{M}\left(0^{-},\tau\right)-i\intop_{t_{0}}^{t}d\bar{t}e^{i\mathbf{h}_{CC}^{eff}\left(\bar{t}-t_{0}\right)}\left(\mathbf{\Sigma}_{CC}^{\urcorner}\star\mathbf{G}_{CC}^{M}\right)_{\left(\bar{t},\tau\right)}\right]\label{eq:Right-GF-via-integral}
\end{equation}
Similarly, 
\begin{equation}
\mathbf{G}_{CC}^{\ulcorner}\left(\tau,t\right)=\left[\mathbf{G}_{CC}^{M}\left(\tau,0^{-}\right)+i\intop_{t_{0}}^{t}d\bar{t}\left(\mathbf{G}_{CC}^{M}\star\mathbf{\Sigma}_{CC}^{\ulcorner}\right)_{\left(\tau,\bar{t}\right)}e^{-i\left(\mathbf{h}_{CC}^{eff}\right)^{\dagger}\left(\bar{t}-t_{0}\right)}\right]e^{i\left(\mathbf{h}_{CC}^{eff}\right)^{\dagger}\left(t-t_{0}\right)}\label{eq:Left-GF-via-integral}
\end{equation}
Note that the obtained expressions contain exponential functions of
the matrix $\mathbf{h}_{CC}^{eff}$. Formally these expressions for
the right and left Green's functions in terms of different components
are identical to those given in \citep{RVLS2013} for the constant
bias switch-on case. Note, however, that this apparent similarity
is misleading as the detailed expressions for the right and left self-energies
are not the same in the two cases.

\subsubsection{Retarded and advanced Green's functions}

It is well known that for the case of the static bias the retarded
and advanced Green's functions of the whole system still depend on
the time difference as in the stationary case. It is crucial for our
derivation, in which the bias applied to each lead is time-dependent,
that this property of these two Green's functions still holds. This
can be shown by first projecting Eq. (\ref{eq:EoM-for-G_CC-1}) on
the retarded subspace of the $(z_{1},z_{2})$ plane and applying the
Langreth rules: 
\begin{equation}
\left(i\frac{d}{dt_{1}}-\mathbf{h}_{CC}\right)\mathbf{G}_{CC}^{r}\left(t_{1},t_{2}\right)=\mathbf{1}_{CC}\delta\left(t_{1}-t_{2}\right)+\left(\mathbf{\Sigma}_{CC}^{r}\cdot\mathbf{G}_{CC}^{r}\right)_{\left(t_{1},t_{2}\right)}\label{eq:EoM-retarded-1}
\end{equation}
Using expression (\ref{eq:self-energy-R}) for the retarded self-energy,
the following equation of motion for the retarded Green's function
$\mathbf{G}_{CC}^{r}\left(t_{1},t_{2}\right)$ is obtained:
\begin{equation}
\left(i\frac{d}{dt_{1}}-\mathbf{h}_{CC}^{eff}\right)\mathbf{G}_{CC}^{r}\left(t_{1},t_{2}\right)=\mathbf{1}_{CC}\delta\left(t_{1}-t_{2}\right)\label{eq:EoM-retarded}
\end{equation}
It immediately follows from this equation that the retarded Green's
function depends only on the time difference. This is a direct consequence
of the fact that the retarded self-energy is proportional to the delta
function. Therefore, one can introduce in the usual way the Fourier
transform of the retarded Green's function, 
\begin{equation}
\mathbf{G}_{CC}^{r}\left(\omega\right)=\left(\omega\mathbf{1}_{CC}-\mathbf{h}_{CC}^{eff}\right)^{-1}\label{eq:Reatarded-GF-in-omega}
\end{equation}
 The matrix $\mathbf{h}_{CC}^{eff}$ is not Hermitian; however, one
can define its left and right eigenvectors which share the same eigenvalues.
Using then the corresponding spectral representation of this matrix
and integrating over $\omega$ in the complex plane, the time representation
of the retarded Green's function is obtained: 
\begin{equation}
\mathbf{G}_{CC}^{r}\left(t_{1},t_{2}\right)=-i\theta\left(t_{1}-t_{2}\right)e^{-i\mathbf{h}{}_{CC}^{eff}\left(t_{1}-t_{2}\right)}\label{eq:Retarded-GF-via-times}
\end{equation}

A similar calculation for the advanced Green's function yields:
\begin{equation}
\mathbf{G}_{CC}^{a}\left(\omega\right)=\left(\omega\mathbf{1}_{CC}-\left(\mathbf{h}_{CC}^{eff}\right)^{\dagger}\right)^{-1}\label{eq:Advanced-GF-in-omega}
\end{equation}
\begin{equation}
\mathbf{G}_{CC}^{a}\left(t_{1},t_{2}\right)=i\theta\left(t_{2}-t_{1}\right)e^{-i\left(\mathbf{h}{}_{CC}^{eff}\right)^{\dagger}\left(t_{1}-t_{2}\right)}\label{eq:Advanced-GF-via-times}
\end{equation}
We see that even in the case of the variable bias the retarded and
advanced functions do not depend on the bias in the WBLA and hence
it is not surprising that they have exactly the same form as in \citep{RVLS2013}.

\subsubsection{The lesser Green's function}

Projecting Eqs. (\ref{eq:EoM-for-G_CC-1}) and (\ref{eq:EoM-for-G_CC-2})
onto the lesser component, using the Langreth rules and the known
expression for the retarded self-energy (\ref{eq:self-energy-R}),
the following equations of motion for the lesser Green's function
are obtained: 
\begin{equation}
\left(i\frac{d}{dt_{1}}-\mathbf{h}^{eff}\right)\mathbf{G}^{<}\left(t_{1},t_{2}\right)=\left(\mathbf{\Sigma}^{<}\cdot\mathbf{G}^{a}+\mathbf{\Sigma}^{\urcorner}\star\mathbf{G}^{\ulcorner}\right)_{\left(t_{1},t_{2}\right)}\label{eq:EoM-for-lesser-1st}
\end{equation}
\begin{equation}
\mathbf{G}_{CC}^{<}\left(t_{1},t_{2}\right)\left[-i\frac{d}{dt_{2}}-\left(\mathbf{h}_{CC}^{eff}\right)^{\dagger}\right]=\left(\mathbf{G}_{CC}^{r}\cdot\mathbf{\Sigma}_{CC}^{<}+\mathbf{G}_{CC}^{\urcorner}\star\mathbf{\Sigma}_{CC}^{\ulcorner}\right)_{\left(t_{1},t_{2}\right)}\label{eq:EoM-for-lesser-2nd}
\end{equation}
where for simplicity of notations we have omitted the $CC$ subscript
in all matrices above as they are defined on the orbitals of the central
region only and hence there should be no ambiguity. We shall retain
this simplified notations in what follows. 

We now look to put the WBLA Kadanoff-Baym equations for the lesser
Green's Functions into a form suitable for the integration. This is
done by introducing the ``tilded'' function $\widetilde{\mathbf{G}}^{<}\left(t_{1},t_{2}\right)$
via the following transformation (similar to the canonical transformation
performed in \citep{RVLS2013} for the equal time case): 
\[
\mathbf{G}_{CC}^{<}\left(t_{1},t_{2}\right)=e^{-i\mathbf{h_{CC}}^{eff}\left(t_{1}-t_{0}\right)}\widetilde{\mathbf{G}}_{CC}^{<}\left(t_{1},t_{2}\right)e^{i\left(\mathbf{h}_{CC}^{eff}\right)^{\dagger}\left(t_{2}-t_{0}\right)}
\]
Differentiating both sides with respect to either $t_{1}$ or $t_{2}$
and using the above equations of motion, we can calculate the partial
derivatives of the tilded function with respect to both times:
\begin{equation}
\frac{\partial\widetilde{\mathbf{G}}_{CC}^{<}\left(t_{1},t_{2}\right)}{\partial t_{1}}=-ie^{i\mathbf{h}_{CC}^{eff}\left(t_{1}-t_{0}\right)}\left(\mathbf{\Sigma}_{CC}^{<}\cdot\mathbf{G}_{CC}^{a}+\mathbf{\Sigma}_{CC}^{\urcorner}\star\mathbf{G}_{CC}^{\ulcorner}\right)_{\left(t_{1},t_{2}\right)}e^{-i\left(\mathbf{h}_{CC}^{eff}\right)^{\dagger}\left(t_{2}-t_{0}\right)}\label{eq:F1}
\end{equation}
\begin{equation}
\frac{\partial\widetilde{\mathbf{G}}_{CC}^{<}\left(t_{1},t_{2}\right)}{\partial t_{2}}=ie^{i\mathbf{h}_{CC}^{eff}\left(t_{1}-t_{0}\right)}\left(\mathbf{G}_{CC}^{r}\cdot\mathbf{\Sigma}_{CC}^{<}+\mathbf{G}_{CC}^{\urcorner}\star\mathbf{\Sigma}_{CC}^{\ulcorner}\right)_{\left(t_{1},t_{2}\right)}e^{-i\left(\mathbf{h}_{CC}^{eff}\right)^{\dagger}\left(t_{2}-t_{0}\right)}\label{eq:F2}
\end{equation}

We explicitly prove in Appendix A that the mixed second derivatives
of the tilded function do not depend on the order of differentiation.
Therefore, one can introduce the exact differential 

\[
d\widetilde{\mathbf{G}}^{<}\left(t_{1},t_{2}\right)=\mathbf{F}^{(1)}\left(t_{1},t_{2}\right)dt_{1}+\mathbf{F}^{(2)}\left(t_{1},t_{2}\right)dt_{2}
\]
so that $\widetilde{\mathbf{G}}^{<}\left(t_{1},t_{2}\right)$ can
be calculated using a line integral taken along an arbitrary path
connecting the initial, $\left(t_{0},t_{0}\right)$, and the final,
$\left(t_{1},t_{2}\right)$, points in the two-dimensional $\left(t_{1},t_{2}\right)$
plane. In Appendix B, details are given on how to perform the line
integral which results in the following final expression for the lesser
Green's function of two times:
\[
\mathbf{G}^{<}\left(t_{1},t_{2}\right)=e^{-i\mathbf{h}^{eff}\left(t_{1}-t_{0}\right)}\int\frac{d\omega}{2\pi}f\left(\omega-\mu\right)\underset{\alpha}{\sum}\left[\mathbf{K}_{\alpha}\left(t_{1},t_{0};\omega\right)\mathbf{\Gamma}_{\alpha}\mathbf{G}^{a}\left(\omega\right)-\mathbf{G}^{r}\left(\omega\right)\mathbf{\Gamma}_{\alpha}\mathbf{K}_{\alpha}^{\dagger}\left(t_{2},t_{0};\omega\right)\right.
\]
\begin{equation}
\left.+i\mathbf{K}_{\alpha}\left(t_{1},t_{0};\omega\right)\mathbf{\Gamma}_{\alpha}\mathbf{K}_{\alpha}^{\dagger}\left(t_{2},t_{0};\omega\right)+i\mathbf{A}_{\alpha}\left(\omega\right)\right]e^{i\left(\mathbf{h}^{eff}\right)^{\dagger}\left(t_{2}-t_{0}\right)}\label{eq:lesser-GF-final}
\end{equation}
Above, we have introduced the spectral function $\mathbf{A}_{\alpha}\left(\omega\right)=\mathbf{G}^{r}\left(\omega\right)\mathbf{\Gamma}_{\alpha}\mathbf{G}^{a}\left(\omega\right)$
for the $\alpha$ lead, and defined the following matrix object:
\begin{equation}
\mathbf{K}_{\alpha}\left(t,t_{0};\omega\right)=\intop_{t_{0}}^{t}d\bar{t}e^{-i\left(\omega\mathbf{1}-\mathbf{h}^{eff}\right)\left(\bar{t}-t_{0}\right)}e^{-i\psi_{\alpha}\left(\bar{t},t_{0}\right)}\label{eq:K-integral}
\end{equation}
which contains all the effects related to the variable bias. 

The general result obtained for the lesser Green's function allows
calculation of the current through any lead which will be done in
the next section. Here we shall mention that another observable of
great interest which can be directly calculated from the Green's function
is the central region density matrix $\mathbf{\rho}\left(t\right)=-i\mathbf{G}^{<}\left(t,t\right)$. 

Note that in \citep{RVLS2013} only the equal time lesser Green's
function was derived as this is sufficient to calculate the current
in WBLA. However, the two-time Green's function may be required for
the calculation of other physical quantities, e.g. for the calculation
of the noise \citep{MRAMLK2014}), and hence we have provided the
general result here.

\subsection{The current}

\subsubsection{Generalized LB formula}

The LB formula for the current in lead $\alpha$ is obtained within
the NEGF formalism as the long-time limit of the quantity $2q\left\langle \frac{d\hat{N}_{\alpha}\left(t\right)}{dt}\right\rangle $
following a static bias switch-on\citep{diventra,RVLS2013}, where
$\hat{N}_{\alpha}\left(t\right)=\underset{k}{\sum}\hat{d}_{k\alpha}^{\dagger}\hat{d}_{k\alpha}$
is the number operator of this lead. Using the equation of motion
for the number operator, it can be shown that, for all times, this
current is given by the following expression \citep{RVLS2013}:

\[
I_{\alpha}\left(t\right)=4qReTr_{C}\left[\mathbf{h}_{C\alpha}\mathbf{G}_{\alpha C}^{<}\left(t,t\right)\right]
\]
One then applies the Langreth theorem to the Dyson equation for $\mathbf{G}_{\alpha C}^{<}\left(t,t\right)$
to obtain: 

\begin{equation}
I_{\alpha}\left(t\right)=-4Re\,\mbox{Tr}_{C}\left[\left(\mathbf{\Sigma}_{\alpha}^{<}\cdot\mathbf{G}^{a}+\mathbf{\Sigma}_{\alpha}^{r}\cdot\mathbf{G}^{<}+\mathbf{\Sigma}_{\alpha}^{\urcorner}\star\mathbf{G}^{\ulcorner}\right)_{\left(t,t\right)}\right]\label{eq:current-Def}
\end{equation}
where the trace is taken with respect to the orbitals of the central
region and we set the charge of an electron to $-1$. This expression
in its right hand side contains the sum of three convolution integrals
defined by Eqs. (\ref{eq:Def-of-dot-product}) and (\ref{eq:Def-of-star-product}).
Let us consider briefly the physical meaning of each term in this
current formula. $\mathbf{\Sigma}_{\alpha}^{<}$ is proportional to
the decoupled lesser Green's function of the $\alpha$ lead, and therefore
to the probability of finding an electron in the leads, whereas $\mathbf{G}^{a}$
is a two-time propagator. So $\mathbf{\Sigma}_{\alpha}^{<}\cdot\mathbf{G}^{a}$
is interpreted as a 'source' term: it gives the current due to electrons
tunneling from lead $\alpha$ into the central region\citep{RVL2013}.
The $\mathbf{\Sigma}_{\alpha}^{r}\cdot\mathbf{G}^{<}$ term, by contrast,
is a `sink' term: it contains $\mathbf{G}^{<}$, which is proportional
to the probability of occupation of the $C$ region, and may therefore
be attributed to the tunneling of electrons out from the central region\citep{JAUHO1994,RVL2013}.
Finally, $\mathbf{\Sigma}_{\alpha}^{\urcorner}\star\mathbf{G}^{\ulcorner}$
is a term containing information on the coupled system prior to the
switch on time $t_{0}$. As such, it vanishes in the partitioned approach,
and the formula (\ref{eq:current-Def}) reduces to the expression
used in Ref. \citep{JAUHO1994}. Within the WBLA, all functions of
two times appearing in these convolution integrals are known, so that
after some lengthy but simple algebra the current evaluates to the
following:
\[
I_{\alpha}(t)=2i\int\frac{d\omega}{2\pi}f\left(\omega-\mu\right)\,\mbox{Tr}_{C}\left\{ \mathbf{\Gamma}_{\alpha}e^{i\left(\omega\mathbf{1}-\mathbf{h}^{eff}\right)\left(t-t_{0}\right)}e^{i\psi_{\alpha}\left(t,t_{0}\right)}\left[\mathbf{G}^{r}\left(\omega\right)-i\mathbf{K}_{\alpha}\left(t,t_{0};\omega\right)\right]\right.
\]
\[
-\mathbf{\Gamma}_{\alpha}\left[\mathbf{G}^{a}\left(\omega\right)+i\mathbf{K}_{\alpha}^{\dagger}\left(t,t_{0};\omega\right)\right]e^{-i\left(\omega\mathbf{1}-\mathbf{h}^{eff}\right)^{\dagger}\left(t-t_{0}\right)}e^{-i\psi_{\alpha}\left(t,t_{0}\right)}+\mathbf{\Gamma}_{\alpha}e^{-i\mathbf{h}^{eff}\left(t-t_{0}\right)}\underset{\alpha^{\prime}}{\sum}\left[i\mathbf{K}_{\alpha^{\prime}}\left(t,t_{0};\omega\right)\mathbf{\Gamma}_{\alpha^{\prime}}\mathbf{K}_{\alpha^{\prime}}^{\dagger}\left(t,t_{0};\omega\right)\right.
\]
\begin{equation}
\left.\left.+\mathbf{K}_{\alpha^{\prime}}\left(t,t_{0};\omega\right)\mathbf{\Gamma}_{\alpha^{\prime}}\mathbf{G}^{a}\left(\omega\right)-\mathbf{G}^{r}\left(\omega\right)\mathbf{\Gamma}_{\alpha^{\prime}}\mathbf{K}_{\alpha^{\prime}}^{\dagger}\left(t,t_{0};\omega\right)+i\mathbf{A}_{\alpha^{\prime}}\left(\omega\right)\right]e^{i\left(\mathbf{h}^{eff}\right)^{\dagger}\left(t-t_{0}\right)}\right\} \label{eq:final-current}
\end{equation}

This expression enables us to model electron transport in response
to the switch-on of an arbitrary time-dependent bias, within the WBLA.

\subsubsection{Recovery of known results: asymptotics}

Without making any assumptions on the shape of the bias, it is instructive
to investigate the long time behaviour of the expression (\ref{eq:final-current}),
which can be done by taking the limit $t_{0}\rightarrow-\infty$,
i.e. we shift the switch-on time to the distant past. We first note
that, as $t_{0}\rightarrow-\infty$, the asymptotic value of the 'preparation'
term,
\[
\mathbf{\Sigma}_{\alpha}^{\urcorner}\star\mathbf{G}^{\ulcorner}\left(t,t\right)=i\int\frac{d\omega}{2\pi}f\left(\omega-\mu\right)\mathbf{\mathbf{\Gamma}}_{\alpha}\mathbf{G}^{a}\left(\omega\right)e^{-i\left(\omega\mathbf{1}-\mathbf{h}_{CC}^{eff}\right)^{\dagger}\left(t-t_{0}\right)}
\]
which contains the matrix function $e^{-i\left(\omega\mathbf{1}-\mathbf{h}_{CC}^{eff}\right)^{\dagger}\left(t-t_{0}\right)}\propto e^{\Gamma t_{0}/2}$,
is vanishing ($\Gamma>0$), so that it has no effect on the steady
state current. To investigate other objects associated with the system
preparation, we consider those terms which result from vertical contour
convolutions in $\mathbf{G}^{<}$ (i.e. terms arising from $\mathcal{G}^{(2)}$
and $\mathcal{G}^{(4)}$ in Appendix B):

\[
\underset{t_{0}\rightarrow-\infty}{\lim}\left[e^{-i\mathbf{h}_{CC}^{eff}\left(t-t_{0}\right)}\left(\mathbf{K}_{\alpha}\left(t,t_{0};\omega\right)\mathbf{\Gamma}_{\alpha}\mathbf{G}^{a}\left(\omega\right)-\mathbf{G}^{r}\left(\omega\right)\mathbf{\Gamma}_{\alpha}\mathbf{K}_{\alpha}^{\dagger}\left(t,t_{0};\omega\right)\right)e^{i\left(\mathbf{h}_{CC}^{eff}\right)^{\dagger}\left(t-t_{0}\right)}\right]=0
\]

\noindent also due to the $e^{\Gamma t_{0}/2}$ term; note that all
times in the integral domain of $\mathbf{K}_{\alpha}$ are less than
or equal to $t$. In addition, the term which arises from the initial
condition for the integral of the lesser Green's function, 
\[
\underset{t_{0}\rightarrow-\infty}{\lim}\left[e^{-i\mathbf{h}_{CC}^{eff}\left(t-t_{0}\right)}\mathbf{A}_{\alpha}\left(\omega\right)e^{i\left(\mathbf{h}_{CC}^{eff}\right)^{\dagger}\left(t-t_{0}\right)}\right]=0
\]
also vanishes in this limit as it also contains a similar exponential
factor of $e^{\Gamma t_{0}}$. The asymptotic behaviour of the current
for an arbitrary time-dependent bias is therefore given by:

\[
\underset{t_{0}\rightarrow-\infty}{\lim}I_{\alpha}(t)=\underset{t_{0}\rightarrow-\infty}{\lim}2\int\frac{d\omega}{2\pi}f\left(\omega-\mu\right)\,\mbox{Tr}_{C}\left\{ \mathbf{\Gamma}_{\alpha}\left(e^{i\left(\omega\mathbf{1}-\mathbf{h}^{eff}\right)\left(t-t_{0}\right)}e^{i\psi_{\alpha}\left(t,t_{0}\right)}\mathbf{K}_{\alpha}\left(t,t_{0};\omega\right)\right.\right.
\]
\begin{equation}
\left.\left.+\mathbf{K}_{\alpha}^{\dagger}\left(t,t_{0};\omega\right)e^{-i\left(\omega\mathbf{1}-\mathbf{h}^{eff}\right)^{\dagger}\left(t-t_{0}\right)}e^{-i\psi_{\alpha}\left(t,t_{0}\right)}-e^{-i\mathbf{h}^{eff}\left(t-t_{0}\right)}\underset{\alpha^{\prime}}{\sum}\mathbf{K}_{\alpha^{\prime}}\left(t,t_{0};\omega\right)\mathbf{\Gamma}_{\alpha^{\prime}}\mathbf{K}_{\alpha^{\prime}}^{\dagger}\left(t,t_{0};\omega\right)e^{i\left(\mathbf{h}^{eff}\right)^{\dagger}\left(t-t_{0}\right)}\right)\right\} \label{eq:asymptote}
\end{equation}

\noindent where the limit remains on the right-hand side to account
for the possibility of decaying terms once the bias has been specified,
and we remark that in this limit the current remains a variable function
of $t$. 

\noindent In the formula (\ref{eq:asymptote}), all terms resulting
from either the equilibrium initial condition or from vertical contour
convolutions have vanished. This tells us that the long-time limit
of the current is independent of the way the system was prepared,
and highlights the advantages of using a contour formalism, in which
all preparation-dependent terms can be easily identified. Furthermore,
with the exception of the initial condition term, every vanishing
term includes a convolution with a left or right self-energy. In the
partitioned approach to the transport problem, these quantities vanish,
as one can see from the definition (\ref{eq:self-energy-Def}) and
$\mathbf{h}_{C\alpha}\left(t_{0}-i\tau\right)=0$. So the formula
(\ref{eq:asymptote}) tells us that, at long times, the partitioned
approach yields the same current as in the partition-free treatment.
If one is interested in the transient regime, however, only the latter
approach will yield a physical current. 

At this point, we can make a precise connection with the results of
Jauho et. al., who considered the switch-on of a time-dependent bias
taking place at $t_{0}=-\infty$ within a partitioned approach\citep{JAUHO1994}.
They solved the Dyson equations on a two-branch contour running from
$-\infty$ to $t$, before returning to $-\infty$ on a lower time
branch, and obtained the following expression for the lesser Green's
function of two times:

\[
\mathbf{G}^{<}\left(t_{1},t_{2}\right)=\underset{-\infty}{\overset{t}{\int\int}}d\bar{t}d\bar{t}'\mathbf{G}^{r}\left(t_{1},\bar{t}\right)\mathbf{\Sigma}^{<}\left(\bar{t},\bar{t}'\right)\mathbf{G}^{a}\left(\bar{t}',t_{2}\right)
\]

\noindent where the functional form of the retarded/advanced Green's
functions is identical to (\ref{eq:Retarded-GF-via-times}), (\ref{eq:Advanced-GF-via-times}).
Working within the WBLA, and neglecting any time-dependence of the
coupling, we can re-write this in the form:

\begin{eqnarray}
\mathbf{G}^{<}\left(t_{1},t_{2}\right) & = & \underset{t_{0}\rightarrow-\infty}{\lim}e^{-i\mathbf{h}_{CC}^{eff}\left(t_{1}-t_{0}\right)}\int\frac{d\omega}{2\pi}f\left(\omega-\mu\right)\underset{\alpha}{\sum}i\mathbf{K}_{\alpha}\left(t_{1},t_{0};\omega\right)\mathbf{\Gamma}_{\alpha}\mathbf{K}_{\alpha}^{\dagger}\left(t_{2},t_{0};\omega\right)e^{i\left(\mathbf{h}_{CC}^{eff}\right)^{\dagger}\left(t_{2}-t_{0}\right)}\label{eq: GLess_Keld}
\end{eqnarray}

\noindent Working on the Keldysh contour used by Jauho et. al., one
neglects imaginary time convolutions when applying the Langreth Rules,
to get instead of (\ref{eq:final-current}) the following expression
for the current:

\begin{equation}
I_{\alpha}\left(t\right)=-4Re\,\mbox{Tr}_{C}\left[\left(\mathbf{\Sigma}_{\alpha}^{<}\cdot\mathbf{G}^{a}+\mathbf{\Sigma}_{\alpha}^{r}\cdot\mathbf{G}^{<}\right)_{\left(t,t\right)}\right]\label{eq:current-Def-Keld}
\end{equation}

\noindent where the real-time convolutions now extend from $-\infty$
up to $t$. Inserting (\ref{eq: GLess_Keld}) into (\ref{eq:current-Def-Keld})
and working within the WBLA, one can easily show that the resulting
formula is given exactly by (\ref{eq:asymptote}). This discussion
clearly shows that the formula derived in \citep{JAUHO1994} is approximate
as it is missing the terms related to the initial preparation of the
system.

\subsubsection{Recovery of known results: static bias}

A second important check on the general formula (\ref{eq:final-current})
is that it reduces to known expressions for the time-dependent current
in the case of a time-independent bias after the switch-on. When the
bias $V_{\alpha}$ is static after the switch on, (\ref{eq:final-current})
can be evaluated purely in terms of the retarded/advanced Green's
functions of the central region, as in this case,

\begin{equation}
\mathbf{K}_{\alpha}\left(t_{1},t_{0};\omega\right)=i\mathbf{G}^{r}\left(\omega+V_{\alpha}\right)\left[e^{-i\left(\left(\omega+V_{\alpha}\right)\mathbf{1}-\mathbf{h}_{CC}^{eff}\right)\left(t_{1}-t_{0}\right)}-\mathbf{1}\right]\label{eq:static-K-1}
\end{equation}
\begin{equation}
\mathbf{K}_{\alpha}^{\dagger}\left(t_{2},t_{0};\omega\right)=-i\left[e^{i\left(\left(\omega+V_{\alpha}\right)\mathbf{1}-\mathbf{h}_{CC}^{eff}\right)^{\dagger}\left(t_{2},t_{0}\right)}-\mathbf{1}\right]\mathbf{G}^{a}\left(\omega+V_{\alpha}\right)\label{eq:static-K-2}
\end{equation}
\textcolor{black}{Substituting these expressions into Eq. (\ref{eq:lesser-GF-final}),
it is found to reduce to the following formula:
\[
\mathbf{G}^{<}\left(t_{1},t_{2}\right)=i\underset{\alpha}{\sum}\frac{d\omega}{2\pi}f\left(\omega-\mu\right)\left[\mathbf{A}_{\alpha}\left(\omega+V_{\alpha}\right)e^{-i\left(\omega+V_{\alpha}\right)\left(t_{1}-t_{2}\right)}\right.
\]
\[
+V_{\alpha}\left(e^{-i\mathbf{h}^{eff}\left(t_{1}-t_{0}\right)}\mathbf{G}^{r}\left(\omega\right)\mathbf{A}_{\alpha}\left(\omega+V_{\alpha}\right)e^{i\left(\omega+V_{\alpha}\right)\left(t_{2}-t_{0}\right)}+e^{-i\left(\omega+V_{\alpha}\right)\left(t_{1}-t_{0}\right)}\mathbf{A}_{\alpha}\left(\omega+V_{\alpha}\right)\mathbf{G}^{a}\left(\omega\right)e^{i\left(\mathbf{h}^{eff}\right)^{\dagger}\left(t_{2}-t_{0}\right)}\right)
\]
\begin{equation}
\left.+V_{\alpha}^{2}e^{-i\mathbf{h}^{eff}\left(t_{1}-t_{0}\right)}\mathbf{G}^{r}\left(\omega\right)\mathbf{A}_{\alpha}\left(\omega+V_{\alpha}\right)\mathbf{G}^{a}\left(\omega\right)e^{i\left(\mathbf{h}^{eff}\right)^{\dagger}\left(t_{2}-t_{0}\right)}\right]\label{eq:static-Gless}
\end{equation}
}

\noindent \textcolor{black}{We remark that this expression is not,
in general, a function of the time difference $\tau\equiv t_{1}-t_{2}$.
This property is satisfied only when both (a) $t_{0}\rightarrow-\infty$
(or $t_{1},t_{2}\rightarrow\infty$) and (b) $\mathbf{h}_{CC}$ commutes
with $\mathbf{\Gamma}$, in which case the Fourier transform of (\ref{eq:static-Gless})
is a function of a single frequency. If one sets $t_{1}=t_{2}$, (\ref{eq:static-Gless})
reduces to the expression published in Refs. \citep{RVLS2013,RVL2013,RVL2014}
for the equal-time case. This can be used to extract the particle
number density of the central region via $\mathbf{\rho}\left(t\right)=-i\mathbf{G}^{<}\left(t,t\right)$.}

\noindent The LB formalism was initially developed for the treatment
of junctions having a static bias placed across them. If one replaces
$\mathbf{K}_{\alpha}$, $\mathbf{K}_{\alpha}^{\dagger}$ with those
given in Eqs. (\ref{eq:static-K-1}) and (\ref{eq:static-K-2}), one
recovers the expression reported in {[}\citep{RVL2013}{]}. \textcolor{black}{If
one then inserts these expressions into (\ref{eq:asymptote}) and
takes the right-hand limit, one obtains a Landauer-type formula for
the multi-terminal junction \citep{RVLS2013}:}

\textcolor{black}{
\begin{equation}
I_{\alpha}^{(LB)}=\underset{t_{0}\rightarrow-\infty}{\lim}I_{\alpha}(t)=2\underset{\alpha'}{\sum}\int\frac{d\omega}{2\pi}\left[f\left(\omega-\mu-V_{\alpha}\right)-f\left(\omega-\mu-V_{\alpha'}\right)\right]\,\mbox{Tr}_{C}\left[\mathbf{T}^{\left(\alpha\alpha'\right)}\left(\omega\right)\mathbf{T}^{\dagger\left(\alpha\alpha'\right)}\left(\omega\right)\right]\label{eq: landauer}
\end{equation}
}

\noindent \textcolor{black}{Here we have defined the transmission
matrix:}

\textcolor{black}{
\begin{eqnarray*}
\mathbf{T}^{\left(\alpha\alpha'\right)}\left(\omega\right) & \equiv & \left[\mathbf{\Gamma}_{\alpha}\right]^{\frac{1}{2}}\mathbf{G}^{r}\left(\omega\right)\left[\mathbf{\Gamma}_{\alpha'}\right]^{\frac{1}{2}}
\end{eqnarray*}
}

\noindent \textcolor{black}{$\mathbf{T}^{\left(\alpha\alpha'\right)}\left(\omega\right)$
is naturally interpreted as the probability amplitude for an electron
to hop from lead $\alpha'$ to lead $\alpha$, so that the current
in lead $\alpha$ is just the sum over all frequencies of the rate
at which electrons in lead $\alpha$ hop into $\alpha'$, minus the
rate at which electrons in lead $\alpha'$ hop into $\alpha$. As
all dependence on $t$ has vanished in the long-time and static bias
limits, these rates are also fixed for all times.}

\section{Results\label{sec:Results}}

To illustrate the general formula (\ref{eq:final-current}) derived
above, we specialize the discussion to that\textcolor{black}{{} of resonant
tunneling through a single-level quantum dot, coupled to just two
leads, L and R. The Hamiltonian matrix $\mathbf{h}_{CC}$ of the central
region is given by the scalar $h=\varepsilon_{0}$, and all other
matrices defined on the central region also become scalars. In addition,
we assume that the two leads are symmetric, i.e. the coupling is the
same: $\Gamma_{L}=\Gamma_{R}=\frac{1}{2}\Gamma$. We choose our zero
of energy to be equal to the chemical potential by setting $\mu=0$.
Thus the intrinsic frequency scales of the system are given by the
difference $\left|\varepsilon_{0}-\mu\right|$ and by $\Gamma$. The
temperature is set to a value which is an order of magnitude smaller
than any other characteristic energy of the problem: $k_{B}T=0.1$.}

We consider a bias in the left lead that is switched on instantaneously
and thereafter sinusoidally varying in time with frequency $\Omega_{L}$
about a fixed value $V_{L}$, and a bias in the right lead that is
set to zero:

\begin{eqnarray}
V_{L}\left(t\right) & = & V_{L}+A_{L}\cos\left(\Omega_{L}t\right)\:,\; V_{R}\left(t\right)=0\:,\; t\geq t_{0}=0\label{eq:bias}
\end{eqnarray}
Either bias is zero at $t<0$. This particular choice of the bias
introduces two new energy scales into the problem: firstly, there
is an intrinsic energy difference of $\left|\mu+V_{L}-\varepsilon_{0}\right|$
between the fixed point of the left lead Fermi level and the dot energy.
In addition, there is a new energy scale set by the driving frequency
$\Omega_{L}$. When $A_{L}\neq0$, this driving frequency causes the
Fermi level of L to vary on a timescale that is slower than the inverse
plasmon frequency $\omega_{p}$ of the leads, so that at each time
step there is a new energy gap which must be traversed for electron
hopping between the quantum dot and the lead L to occur. 

To evaluate the current in the left lead we use a well-known expression
for the generating function for the $n$-th order Bessel functions
of the first kind $J_{n}\left(z\right)$, so that the time integrals
can be calculated analytically (cf. \citep{JAUHO1994}). This puts
the $K_{L}$ time integral (\ref{eq:K-integral}) into the form 
\begin{gather}
K_{L}\left(t,0;\omega\right)=i\underset{n=-\infty}{\overset{\infty}{\sum}}J_{n}\left(\frac{A_{L}}{\Omega_{L}}\right)G^{r}\left(\omega+V_{L}+n\Omega_{L}\right)\left[e^{-i\left(\omega+V_{L}+n\Omega_{L}-h^{eff}\right)t}-1\right]\label{eq: dotKL}
\end{gather}
where $h^{eff}=h-i\Gamma/2$, while the $K_{R}$ integral is simply
\begin{gather}
K_{R}\left(t,0;\omega\right)=iG^{r}\left(\omega\right)\left[e^{-i\left(\omega-h^{eff}\right)t}-1\right]\label{eq: dotKR}
\end{gather}
These are substituted into (\ref{eq:final-current}) to yield a frequency
integral for the time-dependent current, which may then be calculated
numerically, and we shall plot in the following the resulting current
in the left lead, $I_{L}\left(t\right)$, for various combinations
of the parameters of the problem. These plots will also exhibit the
suitably normalized time dependent bias (dotted line) as well as the
prediction of the current calculated using the steady-state (SS) expression
for the current at the instantaneous value of the bias (red line):
\begin{equation}
I_{L}^{(LB)}=\frac{\Gamma^{2}}{2}\int\frac{d\omega}{2\pi}\frac{\left[f\left(\omega-V-\mu\right)-f\left(\omega-\mu\right)\right]}{\left(\omega-\varepsilon_{0}\right)^{2}+\left(\Gamma/2\right)^{2}}\label{eq:SS-LCR}
\end{equation}
The exact time-dependent (TD) current response calculated using our
new formula will be shown by a black line. Comparison of the exact
current with the one calculated using the SS expression is intended
to illustrate the additional effects that can be captured with an
exact TD approach.

\begin{figure}
\subfloat[]{\includegraphics[height=4cm]{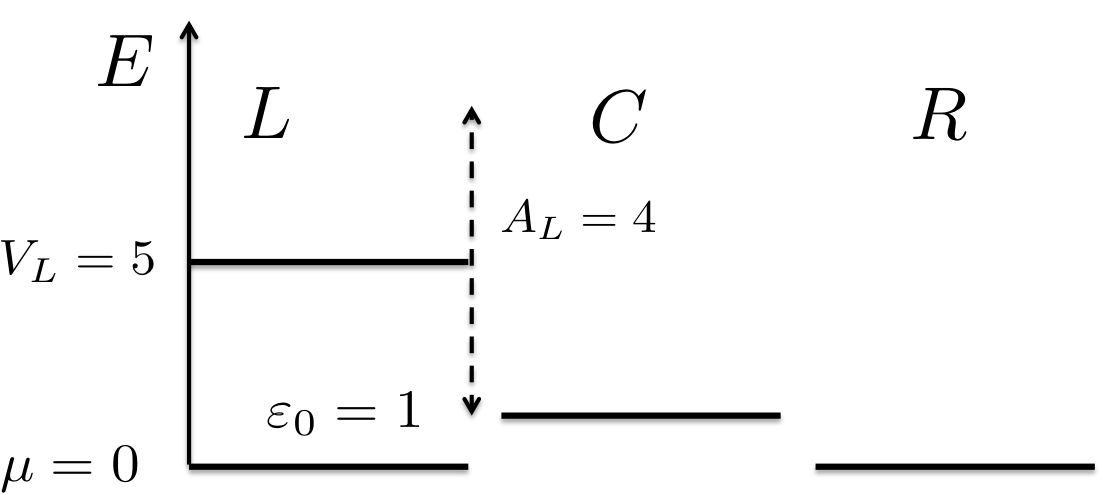}}\subfloat[]{\includegraphics[height=4cm]{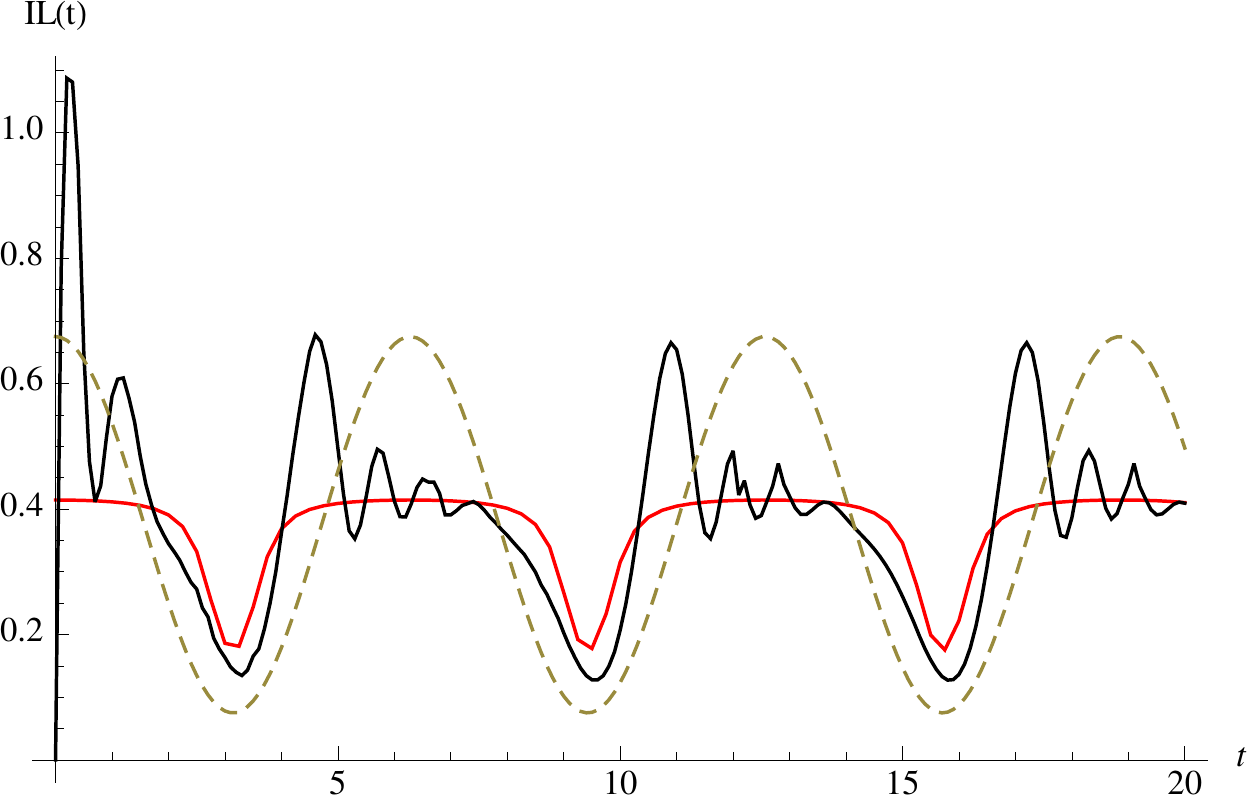}

}

\caption{The basic system we are considering, with parameters $\Gamma=1,\,\varepsilon_{0}=1,\,\mu=0,\, V_{L}=5,\, A_{L}=4,\,\Omega_{L}=1$.
This system is illustrated schematically in (a), and the time-dependent
current through this system shown in (b).\label{fig:The-basic-system}}
\end{figure}

We illustrate the set-up schematically in Fig. \ref{fig:The-basic-system}(a)
for the parametrisation $\varepsilon_{0}=1$, $\mu=0$, $V_{L}=5$
and $A_{L}=4$. In this particular case, the effect of the bias is
to cause the Fermi level of the left lead to move down and 'touch'
the dot energy, before being removed far away from $\varepsilon_{0}$.
\textcolor{black}{The corresponding time-dependence of the current
is displayed in Fig. \ref{fig:The-basic-system}(b) for $\Gamma=1$
and $\Omega_{L}=1$. It has several features in common with all calculated
currents we shall discuss below. }

To understand the SS current, we need only take the zero temperature
limit of the formula (\ref{eq: landauer}) for the two lead case we
are considering. This gives the following expression for the current:

\begin{equation}
\underset{t_{0}\rightarrow-\infty,\, T\rightarrow0}{\lim}I_{L}\left(t\right)=\frac{\Gamma}{2\pi}\left[\arctan\left[\frac{\mu+V_{L}-\varepsilon_{0}}{\Gamma/2}\right]-\arctan\left[\frac{\mu-\varepsilon_{0}}{\Gamma/2}\right]\right]\label{eq:zerotemp}
\end{equation}
This has a roughly linear or ohmic behaviour when the chemical potential
is within $\tfrac{1}{2}\Gamma$ of $\varepsilon_{0}$ and saturates
otherwise, as seen in the red curve in Fig. \ref{fig:The-basic-system}(b).
The curve follows the bias for low bias, but is cut-off at higher
bias.\textcolor{black}{{} }

\textcolor{black}{Our results on the TB current show a rapidly varying
transient that relaxes to a periodic steady-state solution on a timescale
of $2/\Gamma$, a fact which can be predicted from the terms multiplied
by exponential factors $e^{-\Gamma t}$ and $e^{-\Gamma t/2}$ in
the analytic formula for the current we derived (see Appendix C for
the details of this formula). All plots therefore exhibit at least
two effects: the transient reaction to the switch-on event and the
time-varying current due to the persistent bias oscillations. This
much can be seen from the graph in Fig. \ref{fig:The-basic-system}(b),
where for times greater than $2/\Gamma$, the current response is
periodic with a period of $\tau_{L}=2\pi/\Omega_{L}$. In addition
to these two features, we observe a `ringing' oscillation of the current
(first observed in Ref. \citep{JAUHO1994}) occurring within each
period, leading to asymmetry of the current about those voltage peaks
located at multiples of $2\pi$. This `ringing' signal is evidence
of an internal frequency of the system at which resonances in the
current signal occur on a shorter timescale than $\tau_{L}$. It consists
of a series of peaks which are damped consecutively before tending
to flatten, and which then agree for a short time with the steady-state
formula. We also observe that, although the steady-state current is
always in phase with the voltage, the minimum of the exact current
is delayed with respect to the voltage.}

\begin{figure}
\subfloat[$\varepsilon_{0}=2$]{\includegraphics[height=4cm]{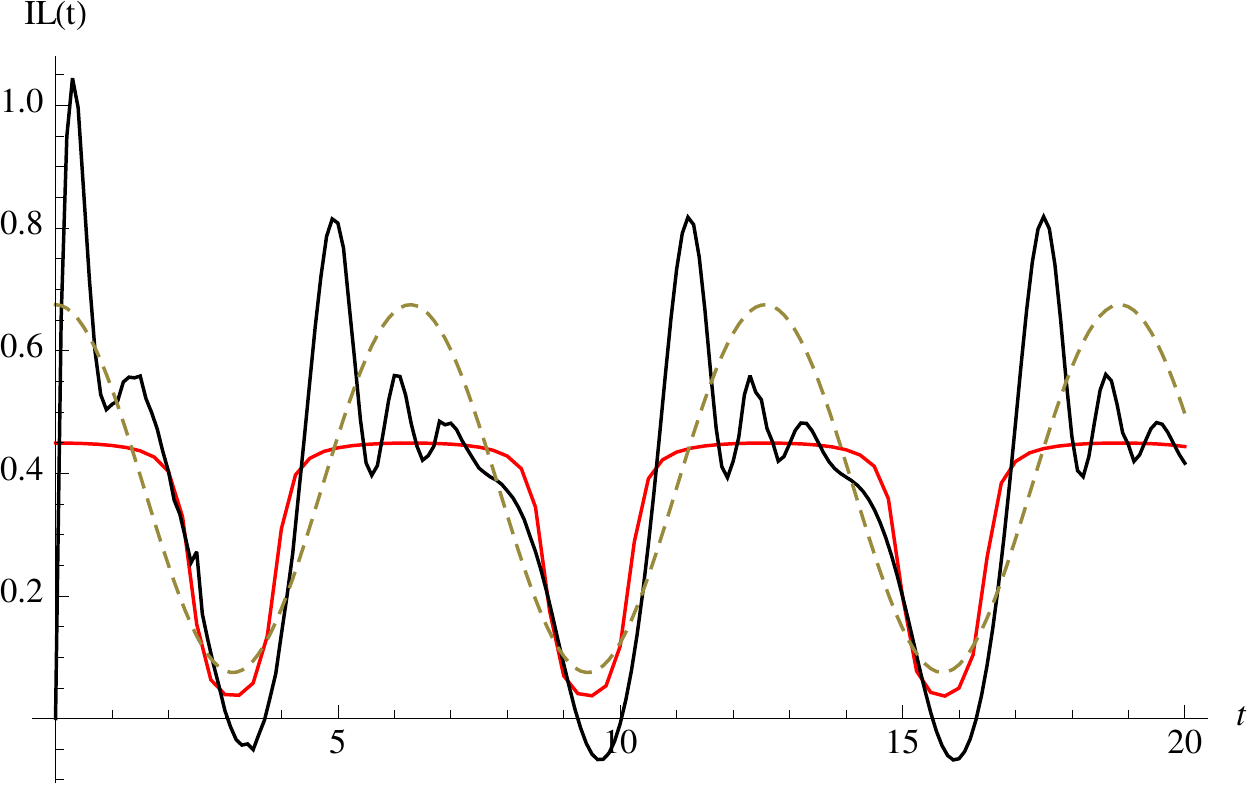}

}\subfloat[$\varepsilon_{0}=3$]{\includegraphics[height=4cm]{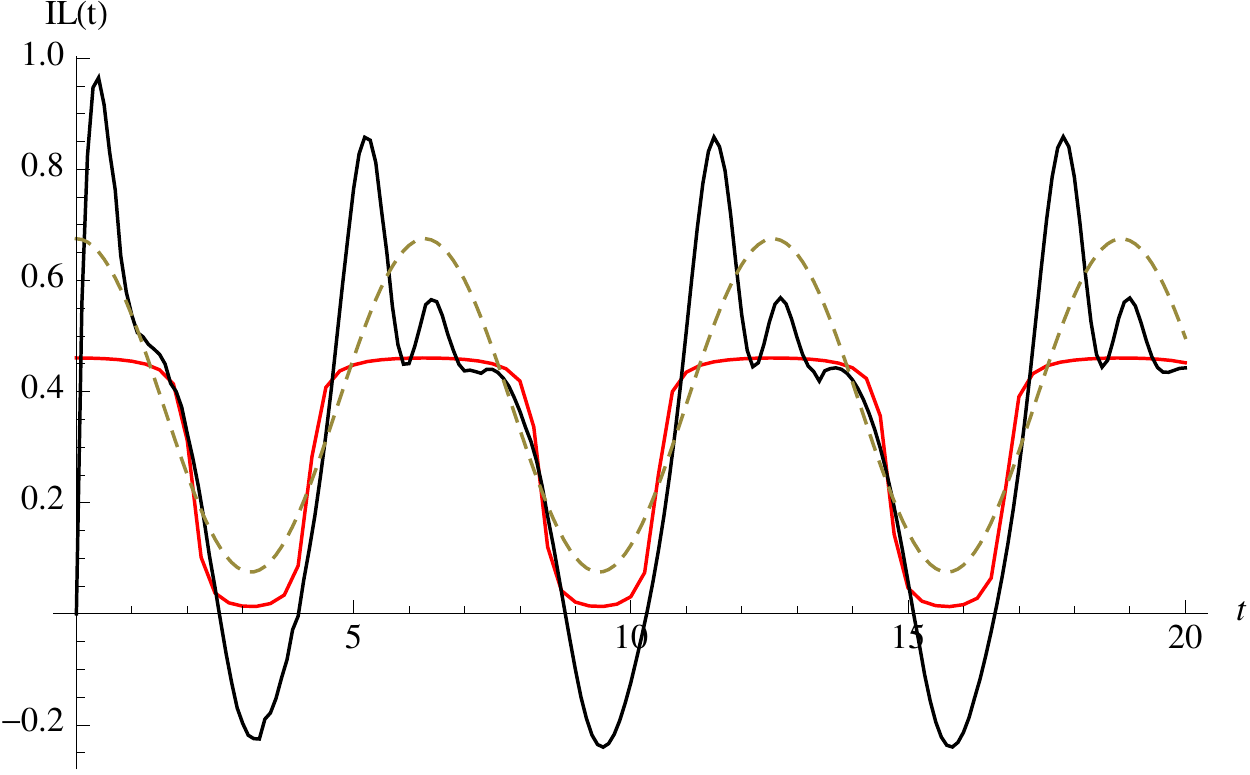}

}

\subfloat[$\varepsilon_{0}=4$]{\includegraphics[height=4cm]{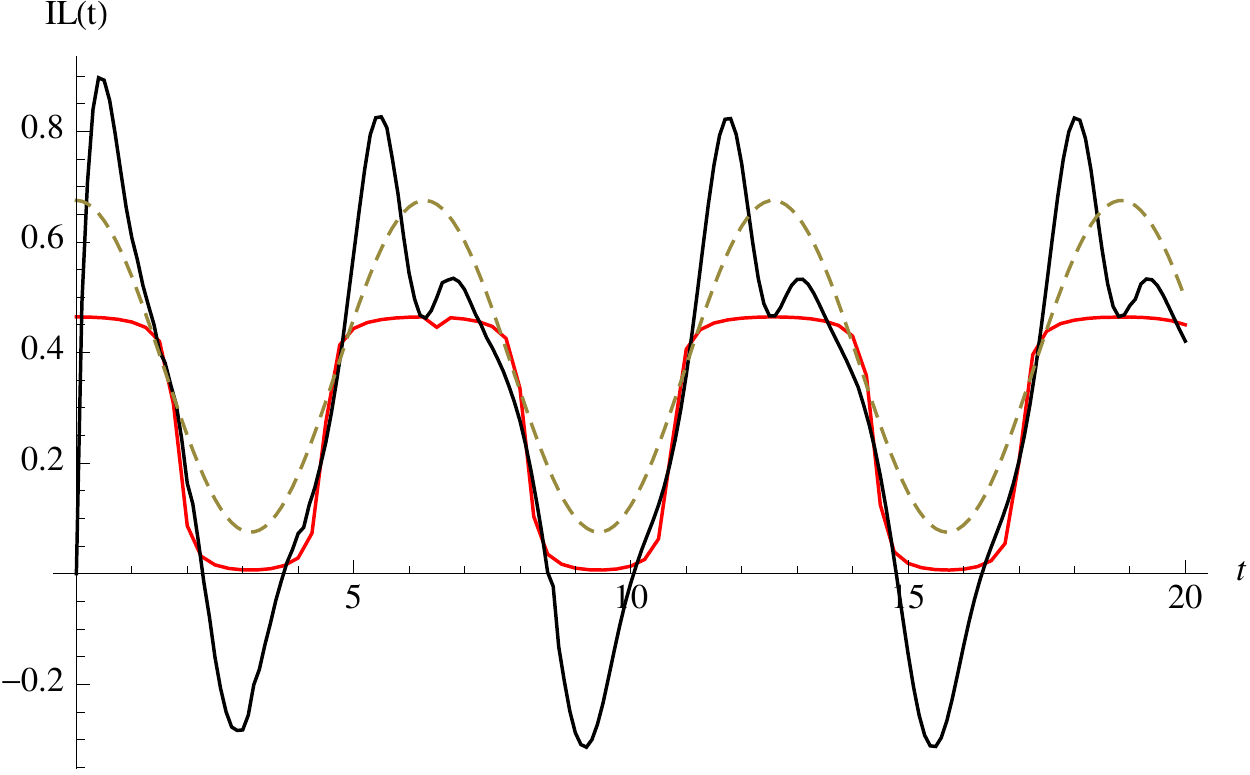}

}\subfloat[$\varepsilon_{0}=5$]{\includegraphics[height=4cm]{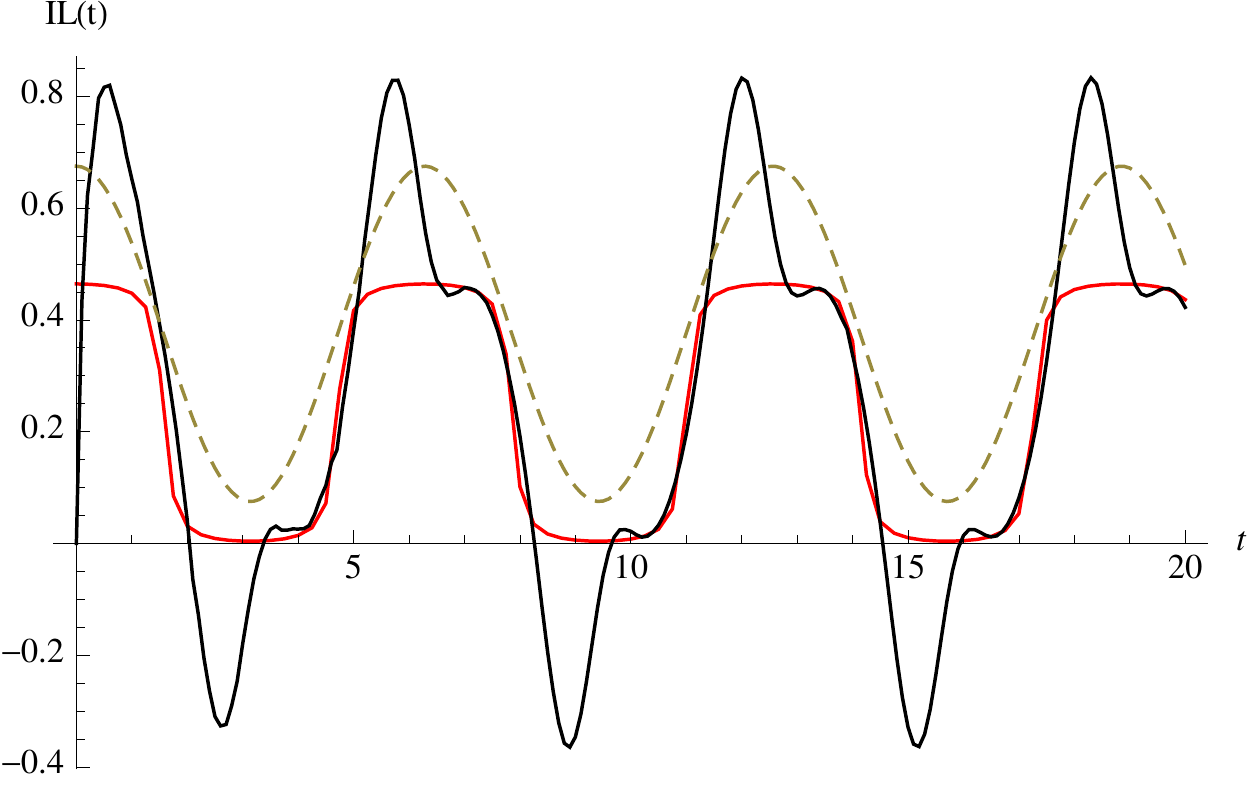}

}

\caption{\label{fig: dotvary}The effect of varying the dot energy. As $\varepsilon_{0}$
is moved closer to $V_{L}$, the intrinsic frequency of the 'ringing'
oscillations is increased. When these two energies are aligned, the
current response is symmetric about $I_{L}\sim0.225$, because the
movement of the left Fermi level is symmetric around $\varepsilon_{0}$.
\textcolor{green}{\label{fig:The-effect-of-dot-level}}}
\end{figure}

Let us first consider the effect on this system of varying the dot
energy level. This is done in steps of unit energy, and the resulting
plots for the current as a function of time are shown in Fig. \ref{fig:The-effect-of-dot-level}(a)-(d).
It is clear from these plots that the frequency of the `ringing' oscillations
of the TD current is determined by the position of the dot relative
to the two Fermi levels. In particular, it seems that the frequency
of these oscillations is roughly equal to the energy gap $\left|\mu+V_{L}-\varepsilon_{0}\right|$,
as each unit increase in $\varepsilon_{0}$ tends to flatten or remove
a peak from the `ringing' transient. The dependence of this effect
on the left gap alone can be checked by moving $V_{L}$ and $\varepsilon_{0}$
by the same amount; indeed our calculations demonstrate that this
leaves the number of `ringing' peaks unchanged. When the gap $\left|\mu+V_{L}-\varepsilon_{0}\right|$
is equal to zero, Fig. \ref{fig:The-effect-of-dot-level}(d), the
`ringing' ceases to be significant, although an asymmetric peak after
the minimum of the bias is still present, as discussed above. This
is not surprising, because when $\varepsilon_{0}=V_{L}$ the time-dependent
shift in the Fermi level of the left lead is symmetric about $\varepsilon_{0}$.
Physically, we interpret the `ringing' current signal as an outcome
of competing intrinsic timescales: the timescale of tunnelling from
the left lead onto the dot is smaller than the typical lifetime of
an electron in the dot state, given by $\tau_{el}\sim2/\Gamma$. It
is apparent from the Figs. \ref{fig:The-effect-of-dot-level}(a)-(d)
that the possibility for a negative current is also determined by
the position of the dot energy relative to the left Fermi level, and
that the minima of the current become increasingly negative as $\varepsilon_{0}$
is brought closer to $V_{L}$.

\begin{figure}
\subfloat[$\Omega_{L}=0.1$. ]{\includegraphics[height=5cm]{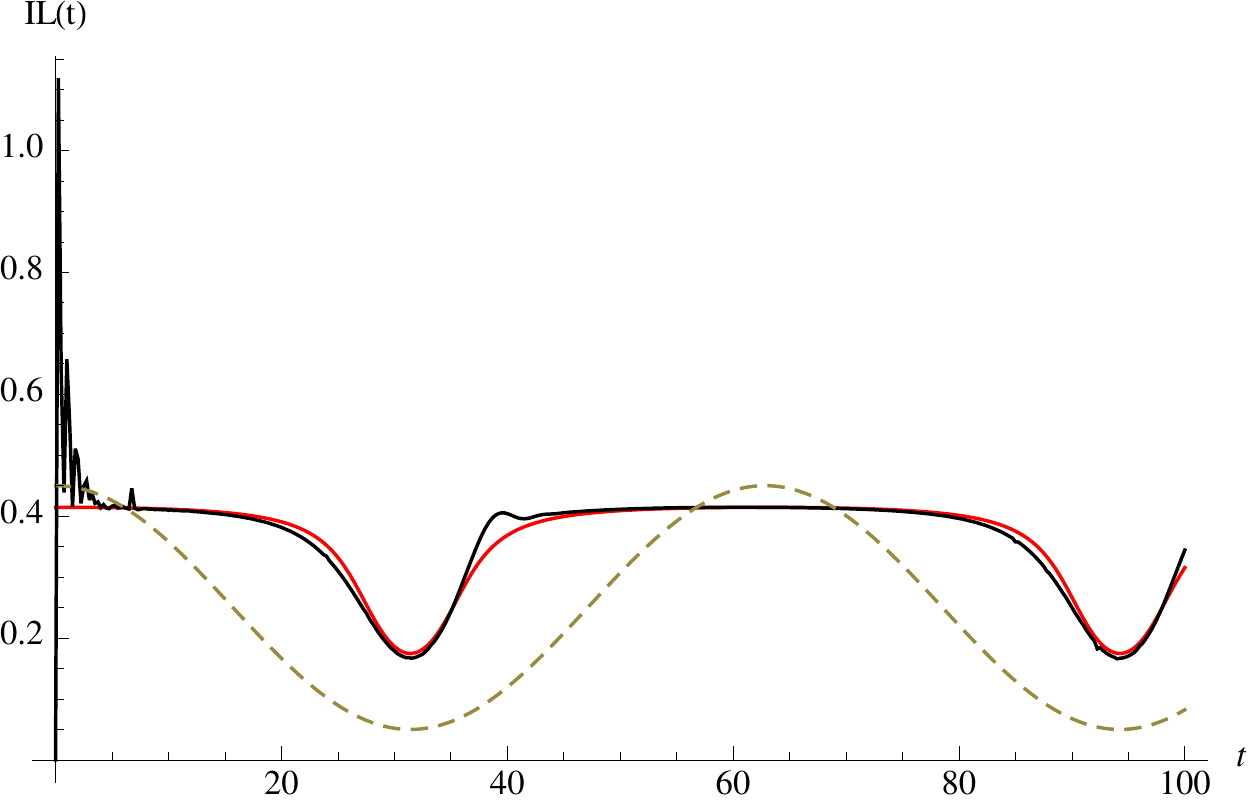}}\subfloat[$\Omega_{L}=10$. ]{\includegraphics[height=5cm]{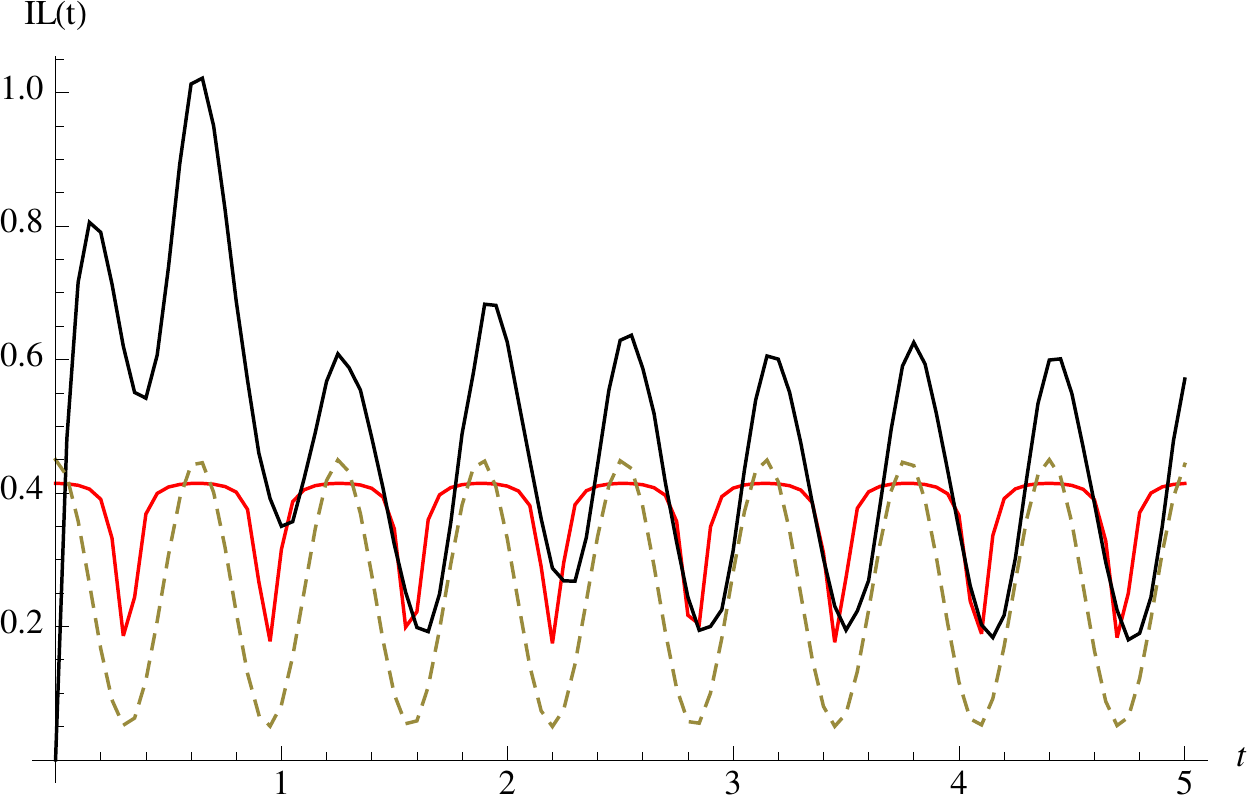}

}\caption{\label{fig:The-effect-of freq}The effect of varying the driving frequency,
with all other parameters as in Fig. \ref{fig:The-basic-system}(b).
In (a), we see that our formula gives a good agreement with the steady-state
LB formula, as this is tending towards the adiabatic limit. In (b),
the high driving frequency causes the current to be phase-delayed,
and also to strongly disagree with the LB formula in amplitude.}
\end{figure}

Next, we consider the effects on the current response of varying the
driving frequency. One would intuitively expect that, as the driving
frequency of the bias is lowered, the agreement between our TD formula
and the LB formula for the SS current becomes better as the bias varies
more slowly, suggesting that the LB formula derived for the case of
a static bias would become appropriate. In Fig. \ref{fig:The-effect-of freq},
all parameters, except for the bias frequency, are the same as in
Fig. \ref{fig:The-basic-system}(b). In (a) the driving frequency
of the bias was scaled down to $\Omega_{L}=0.1$. One can see that
in this adiabatic limit there is excellent agreement with the steady-state
formula, with the exception of a quickly oscillating transient following
$t_{0}$, and a very quickly suppressed single `ringing' peak. In
Fig. \ref{fig:The-effect-of freq}(b), the driving frequency is scaled
up to $\Omega_{L}=10$, with the effect that the current response
deviates significantly from that given by the LB formula. Interestingly,
in this high frequency case it takes around 4 to 5 bias cycles for
the current to stabilise into steady oscillations. In addition, a
phase-shift of the current response with respect to the steady-state
value is clearly visible. During the transient regime, the amplitude
of the exact current disagrees with the steady state current at all
times. The amplitude of the exact current at the bias peaks is much
greater than the amplitude of the steady state current in the long-time
regime.

\begin{figure}
\subfloat[$\Gamma=0.2$]{\includegraphics[height=4cm]{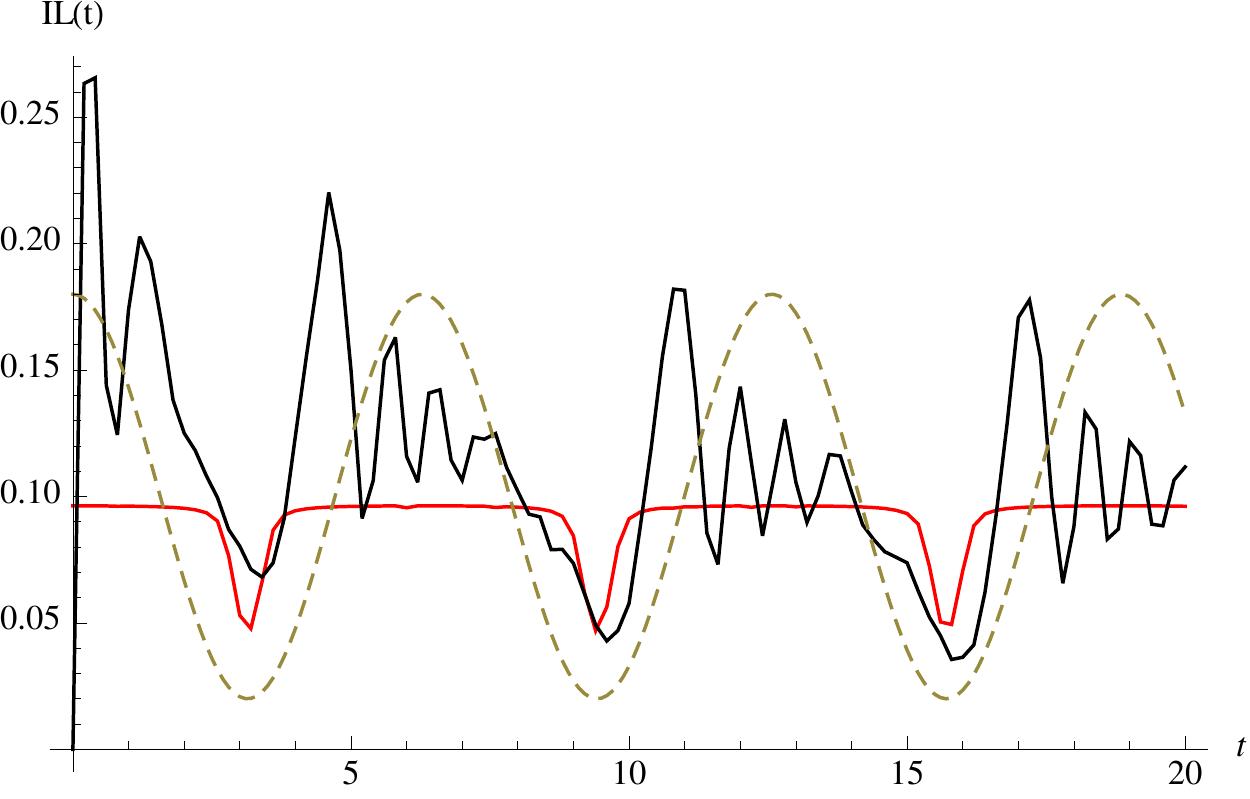}}\subfloat[$\Gamma=5.0$]{\includegraphics[height=5cm]{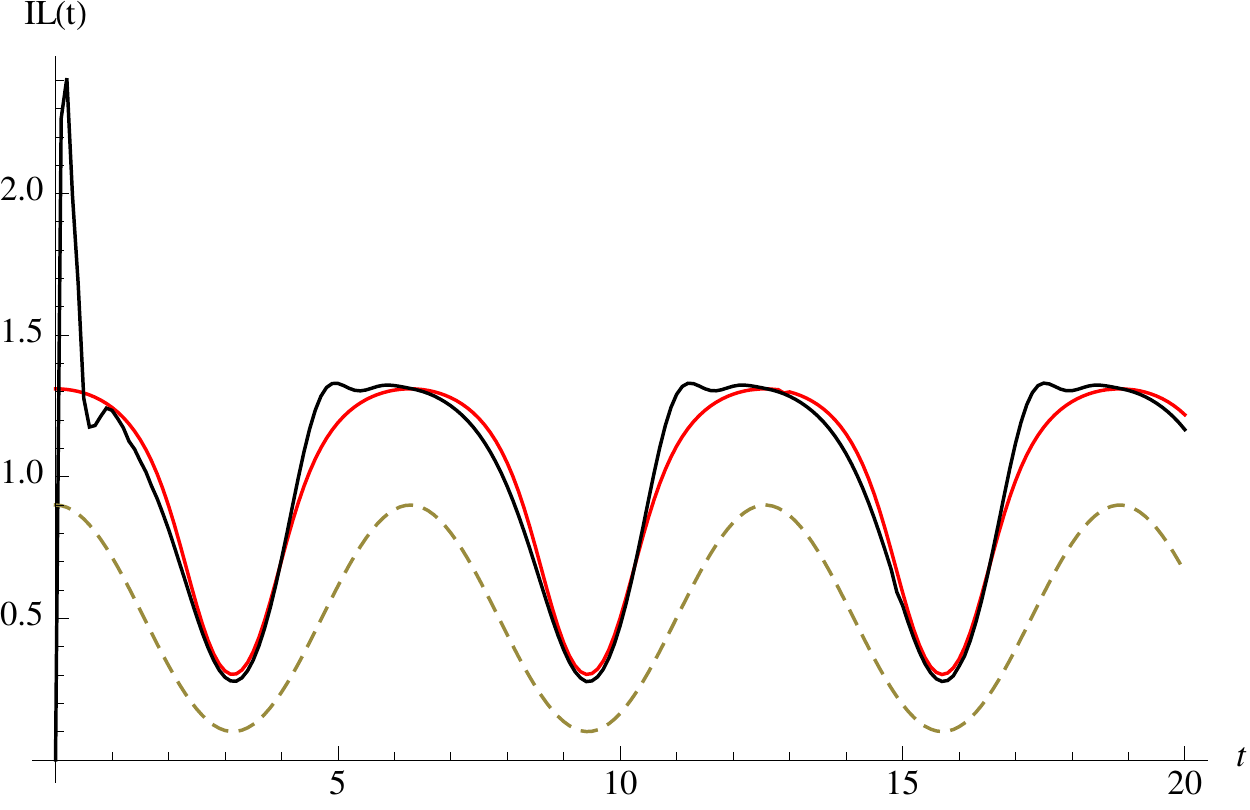}

}

\caption{We exhibit the effects of varying the level width, where all other
parameters as in Fig. \ref{fig:The-basic-system}(b). We find that
$\Gamma$ tends to suppress any non-adiabatic features as it increases
in magnitude.\label{fig:The-level-width-effect}}
\end{figure}

In Fig. \ref{fig:The-level-width-effect} we consider the effect of
varying the level width $\Gamma$, where all other parameters are
fixed at the same values as in Fig. \ref{fig:The-basic-system}(b).
As discussed above, this quantity fixes the timescale over which the
transient current decays. We display the effects of setting $\Gamma=0.2$
in panel (a). We see that this has the effect of prolonging the transient
and of accentuating the `ringing' peaks in the TD current, without
adding any new peaks to this part of the signal. In Fig. \ref{fig:The-level-width-effect}(b)
the level width $\Gamma=$5.0; we see that for $t>2.5=\tau_{el}$,
the TD formula is periodic, but it is also interesting to note that
the `ringing' oscillations are greatly suppressed, in a manner which
is qualitatively similar to the suppression observed above for an
adiabatic driving frequency, Fig. \ref{fig:The-effect-of freq}(a).
Interpreting $\Gamma$ as the frequency with which electrons tunnel
off the dot, we see that in this case it is greater than any other
typical energy gap in the problem, so that the time scale with which
electrons tunnel onto the dot is greater than that with which they
escape.

\begin{figure}
\includegraphics[height=5cm]{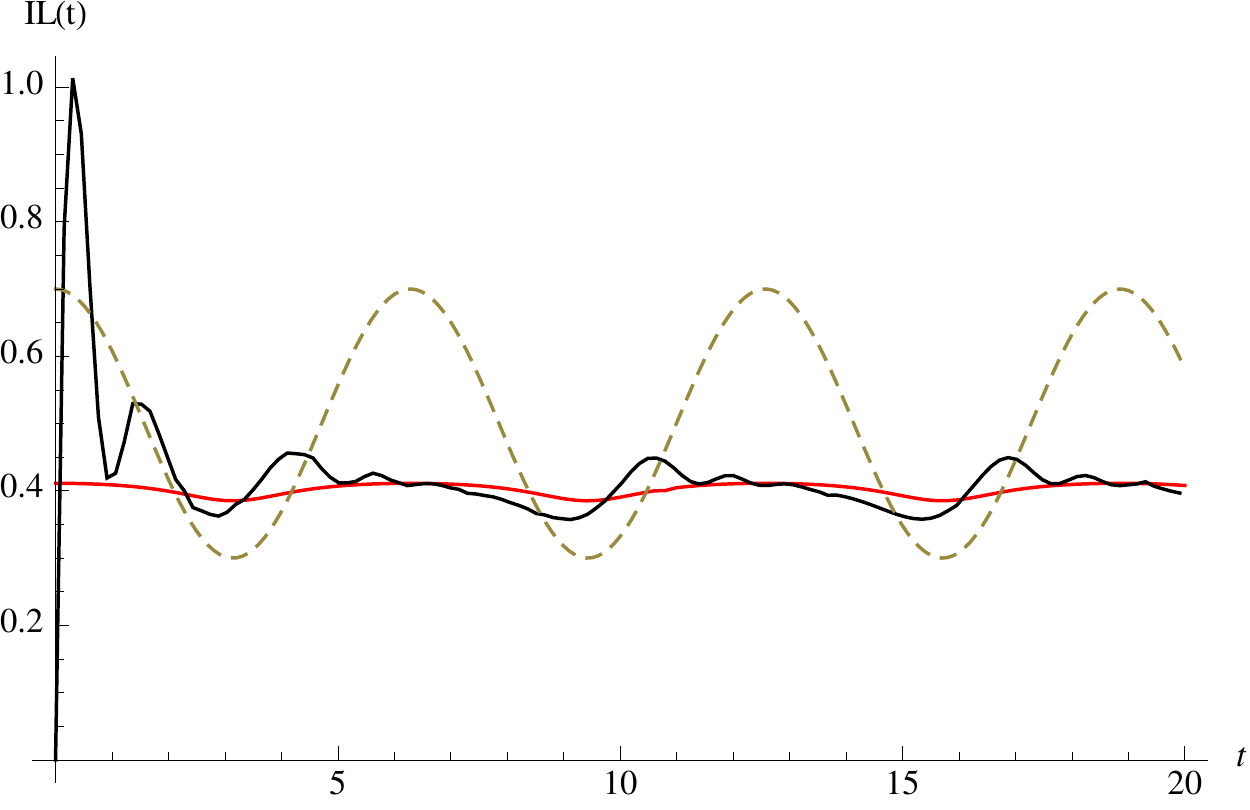}

\caption{Effect on the current response of reducing the bias amplitude from
$A_{L}=4$ (as in Fig. \ref{fig:The-basic-system}(b)) to $A_{L}=2$.
\label{fig:Efect-on-the-amplitude}}
\end{figure}

We also investigate the effect of decreasing the amplitude $A_{L}$
of the bias oscillations about the fixed point $V_{L}$, and the result
for $A_{L}=2$ is shown in Fig. \ref{fig:Efect-on-the-amplitude}.
The main effects observed in Fig. \ref{fig:The-basic-system}(b) are
present, but as the left Fermi level never comes all the way down
to `touch' $\varepsilon_{0}$, there is relatively little `ringing'
in the TD current in addition to the SS behaviour. In the temperature
regime we are considering, Eq. (\ref{eq:zerotemp}) is in close agreement
with the SS formula at all values of the time-dependent bias inserted
into $V_{L}$. Large variations of the bias bring it closer to $\varepsilon_{0}$,
where the arctan function varies most rapidly. For an oscillation
amplitude that simply moves the bias by a small amount around a value
of $V_{L}$ which is much larger than $\varepsilon_{0}$, the SS current
will move up and down an asymptote of the arctan function and remain
almost static, becoming completely static as $A_{L}\rightarrow0$.

\begin{figure}
\includegraphics[height=4cm]{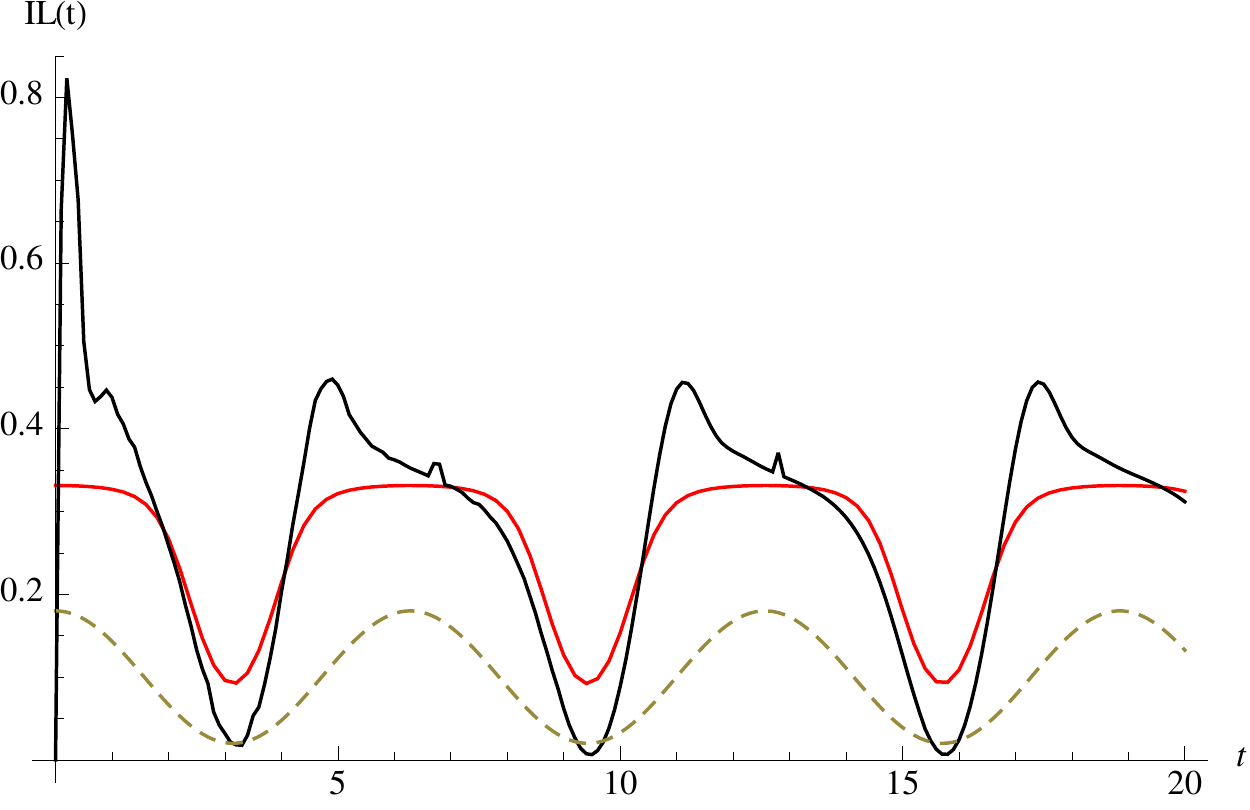}

\caption{\label{fig:high-T}The high temperature behaviour of the current is
investigated. Here we plot the TD current for $k_{B}T=1.0$, and all
other parameters are as in Fig. \ref{fig:The-basic-system}(b).}
\end{figure}

Finally, we can see in Fig. \ref{fig:high-T} that increasing the
system temperature so that $k_{B}T=1.0$ has the effect of suppressing
the `ringing' peaks on the current signal. Increasing the temperature
thus has a qualitatively similar effect to an increase in $\Gamma$.


\section{Conclusions\label{sec:Conclusions}}

In this paper, we have considered, within a tight binding model, electron
conduction in a multi-terminal systrem following the switch-on of
a time dependent bias. Our formalism, which mostly relies on the method
developed in Refs. \citep{StefALm2004,RVL2009,RVL2013,RVLS2013,RVL2014},
is an example of a partition-free approach, whereby the whole system
is fully coupled prior to the bias switch on. 

We showed that within the WBLA it is possible to solve the corresponding
Kadanoff-Baym equations for all components of the Non-Equilibrium
Green's Functions and derive a closed expression for the current in
such a system. Our formula for the current includes both transient
effects due to the switch-on and the current time variation due to
the time-dependent bias in each lead, subsequent to the switch-on.
We showed that our formula reduces to a number of known results previously
obtained: (i) by taking the switch on time $t_{0}$ to $-\infty$,
our formula coincides with the result of Jauho \textit{et. al.} \citep{JAUHO1994}
where a partitioned aproach was used; (ii) assuming the bias is static
after the switch on, we recover the result of Stefanucci \emph{et
al} \citep{StefALm2004,RVLS2013}. 

Moreover, it was also possible to state the conditions under which
the partitioned and partition-free approaches would yield identical
results within the WBLA, and to show that the long-time limit of the
time-dependent current satisfies these conditions. Therefore, as expected
\citep{StefALm2004,RVLS2013}, in the long time limit the expression
for the current due to a constant bias approaches the well-known Landauer
steady-state result. 

The analytical result for the current we have obtained enabled us
to isolate terms in the current which are associated with the system
preparation, and to study the long-time behaviour of the current.
Note that in this case the bias is still time dependent, but all effects
of the switch on have been eliminated.

Finally we have applied the formalism developed to the case of an
AC bias placed across a \textcolor{black}{single-level resonant tunneling
device, with results that reduce to the steady-state LB formalism
in the case where the driving field is varied adiabatically slowly.
We have also looked at effects of the position of the dot energy level
with respect to the Fermi levels of the unbiased leads and showed
that the number of `ringing' oscillations in the current appearing
at the beginning of each bias peak is directly related to the position
of the dot level. Dependence of the current on the oscillation amplitude
of the bias, the temperature, and the coupling between the leads and
the central system (the `level width') has also been analysed.}

We anticipate that this formalism will be applicable to the study
of TD conductance and current fluctuations through magnetic systems
describable by one-particle Hamiltonians (including those extracted
from self-consistent density-functional approaches).

\paragraph{Acknowledgements}

Michael Ridley was supported through a studentship in the Centre for
Doctoral Training on Theory and Simulation of Materials at Imperial
College funded by the Engineering and Physical Sciences Research Council
under grant number EP/G036888/1.

\section*{Appendix A - Path-independence}

It is well-known from mathematics that a necessary and sufficient
condition for the path-independence of a line integral $\int_{L}g(x,y)dx+h(x,y)dy$
on the $(x,y)$ plane is $\partial g/\partial y=\partial h/\partial x$.
This is equivalent to the statement that $df=gdx+hdy$ is an exact
differential of some function $f\left(x,y\right)$. This condition
corresponds to the symmetry of the second order mixed derivatives
$\frac{\partial^{2}f}{\partial x\partial y}=\frac{\partial^{2}f}{\partial y\partial x}$
of this $f\left(x,y\right)$.

Green's functions are defined as functions of two contour times, i.e.
the domain for integration of the Green's function is the set of points
$\left\{ \left\langle z_{1},z_{2}\right\rangle :z_{1}\in\gamma\wedge z_{2}\in\gamma\right\} $.
We shall verify here that the mixed second derivatives of the Green's
function $\mathbf{G}\left(z_{1},z_{2}\right)$ are symmetric. We can
write using the corresponding equations of motion (all matrices are
given on the subspace of the central region and the subscript $CC$
is omitted):
\[
\frac{\partial\mathbf{G}\left(z_{1},z_{2}\right)}{\partial z_{1}}=-i\left[\mathbf{h}\left(z_{1}\right)\mathbf{G}\left(z_{1},z_{2}\right)+\mathbf{1}\delta\left(z_{1},z_{2}\right)+\int_{\gamma}d\bar{z}\mathbf{\Sigma}\left(z_{1},\bar{z}\right)\mathbf{G}\left(\bar{z},z_{2}\right)\right]
\]
\[
\frac{\partial\mathbf{G}\left(z_{1},z_{2}\right)}{\partial z_{2}}=i\left[\mathbf{G}\left(z_{1},z_{2}\right)\mathbf{h}\left(z_{2}\right)+\mathbf{1}\delta\left(z_{1},z_{2}\right)+\int_{\gamma}d\bar{z}\mathbf{G}\left(z_{1},\bar{z}\right)\mathbf{\Sigma}\left(\bar{z},z_{2}\right)\right]
\]

From the first of these:

\[
\frac{\partial^{2}\mathbf{G}\left(z_{1},z_{2}\right)}{\partial z_{2}\partial z_{1}}=-i\left[\mathbf{h}\left(z_{1}\right)\frac{\partial\mathbf{G}\left(z_{1},z_{2}\right)}{\partial z_{2}}+\mathbf{1}\frac{\partial\delta\left(z_{1},z_{2}\right)}{\partial z_{2}}+\int_{\gamma}d\bar{z}\mathbf{\Sigma}\left(z_{1},\bar{z}\right)\frac{\partial\mathbf{G}\left(\bar{z},z_{2}\right)}{\partial z_{2}}\right]
\]
while from the second:

\[
\frac{\partial^{2}\mathbf{G}\left(z_{1},z_{2}\right)}{\partial z_{1}\partial z_{2}}=i\left[\frac{\partial\mathbf{G}\left(z_{1},z_{2}\right)}{\partial z_{1}}\mathbf{h}\left(z_{2}\right)+\mathbf{1}\frac{\partial\delta\left(z_{1},z_{2}\right)}{\partial z_{1}}+\int_{\gamma}d\bar{z}\frac{\partial\mathbf{G}\left(z_{1},\bar{z}\right)}{\partial z_{1}}\mathbf{\Sigma}\left(\bar{z},z_{2}\right)\right]
\]
Note that due to the order in which the Green's functions and self-energy
appear to the right hand side of the derivatives, one never needs
to consider derivatives of the self-energy. Noting next that $\frac{\partial\delta\left(z_{1},z_{2}\right)}{\partial z_{2}}=-\frac{\partial\delta\left(z_{1},z_{2}\right)}{\partial z_{1}}$,
and making use of the first derivatives written above, we obtain that
$\frac{\partial^{2}\hat{G}\left(z_{1},z_{2}\right)}{\partial z_{2}\partial z_{1}}=\frac{\partial^{2}\hat{G}\left(z_{1},z_{2}\right)}{\partial z_{1}\partial z_{2}}$.

Given this identity, we can take the lesser part of the Green's function
by setting $z_{1}=t_{1}\in C_{-}$ and $z_{2}=t_{2}\in C_{+}$, which
results in the required symmetry of the second order mixed derivatives
for the $CC-$component of the lesser Green's function. This fact
can be used to demonstrate that the same property is satisfied by
the object 
\[
\widetilde{\mathbf{G}}^{<}\left(t_{1},t_{2}\right)\equiv e^{\mathbf{B}\left(t_{1}-t_{0}\right)}\mathbf{G}^{<}\left(t_{1},t_{2}\right)e^{\mathbf{B}^{\dagger}\left(t_{2}-t_{0}\right)}
\]
where $\mathbf{B}$ is a square matrix, and therefore path independence
of the line integral 
\begin{equation}
\int_{L}d\widetilde{\mathbf{G}}^{<}=\int_{L}\frac{\partial\widetilde{\mathbf{G}}^{<}}{\partial t_{1}}dt_{1}+\frac{\partial\widetilde{\mathbf{G}}^{<}}{\partial t_{2}}dt_{2}\label{eq:line-integral}
\end{equation}
 is assured, as required.

\section*{Appendix B - Details of the line integration}

The tilded Green's function is calculated using the line integral
(\ref{eq:line-integral}) taken along a particular path $\left(t_{0-},t_{0+}\right)\longrightarrow\left(t_{1},t_{0+}\right)\longrightarrow\left(t_{1},t_{2}\right)$
on the $\left(t_{1},t_{2}\right)$ plane: 

\[
\widetilde{\mathbf{G}}^{<}\left(t_{1},t_{2}\right)=\int_{L}\mathbf{F}^{(1)}\left(t_{1},t_{2}\right)dt_{1}+\mathbf{F}^{(2)}\left(t_{1},t_{2}\right)dt_{2}
\]
\[
=\widetilde{\mathbf{G}}^{<}\left(t_{0-},t_{0+}\right)+\int_{t_{0}}^{t_{1}}d\overline{t}\,\mathbf{F}^{(1)}\left(\overline{t},t_{0+}\right)+\int_{t_{0}}^{t_{2}}d\overline{t}\,\mathbf{F}^{(1)}\left(t_{1},\overline{t}\right)
\]
where the matrices $\mathbf{F}^{(1)}$ and $\mathbf{F}^{(2)}$ for
both times are given by first derivatives of the tilded Green's function,
Eqs. (\ref{eq:F1}) and (\ref{eq:F2}). Here $t_{0-}=t_{0}+i0$ is
the time $t_{0}$ on the upper horizontal part of the contour, while
$t_{0+}=t_{0}-i0$ is the later time lying on the lower horizontal
track of $\gamma$. This guarantees the correct time ordering for
the lesser function. Next we use the fact that the Matsubara Green's
function provides the boundary conditions for the lesser Green's function
at the special point $\left(t_{0-},t_{0+}\right)$: 
\[
\widetilde{\mathbf{G}}^{<}\left(t_{0-},t_{0+}\right)=\mathbf{G}^{<}\left(t_{0-},t_{0+}\right)=\mathbf{G}^{M}\left(0,0^{+}\right)
\]
 This is a way of incorporating information on the system preparation
into a description of its dynamics. Hence, one can write:
\[
\widetilde{\mathbf{G}}\left(t_{1},t_{2}\right)-\mathbf{G}^{M}\left(0,0^{+}\right)=-i\intop_{t_{0}}^{t_{1}}d\bar{t}\, e^{i\mathbf{h}^{eff}\left(\bar{t}-t_{0}\right)}\left[\left(\mathbf{\Sigma}^{<}\cdot\mathbf{G}^{a}\right)+\left(\mathbf{\Sigma}^{\urcorner}\star\mathbf{G}^{\ulcorner}\right)\right]_{\left(\bar{t},t_{0+}\right)}
\]
\begin{equation}
+i\intop_{t_{0}}^{t_{2}}d\bar{t}\, e^{i\mathbf{h}^{eff}\left(t_{1}-t_{0}\right)}\left[\left(\mathbf{G}^{r}\cdot\mathbf{\Sigma}^{<}\right)+\left(\mathbf{G}^{\urcorner}\star\mathbf{\Sigma}^{\ulcorner}\right)\right]_{\left(t_{1},\bar{t}\right)}e^{-i\left(\mathbf{h}^{eff}\right)^{\dagger}\left(\bar{t}-t_{0}\right)}\equiv\sum_{i=1}^{4}\mathbf{\mathcal{G}}^{(i)}\label{eq:long_G-tilde}
\end{equation}

There are four terms here, $\mathcal{G}^{(i)}$ ($i=1,\ldots,4)$,
two coming from each integral, and we shall evaluate them one at a
time. As some parts of the calculation are similar to the one reported
in \citep{RVLS2013}, we only briefly state the main steps here. It
is easily seen that the first term $\mathcal{G}^{(1)}$ in the first
integral in the right hand side above is zero since
\[
\left(\mathbf{\Sigma}^{<}\cdot\mathbf{G}^{a}\right)_{\left(\bar{t},t_{0+}\right)}=\int_{t_{0}}^{\infty}dt^{\prime}\,\mathbf{\Sigma}^{<}\left(\overline{t},t^{\prime}\right)\mathbf{G}^{a}\left(t^{\prime},t_{0+}\right)
\]
contains the advanced Green's function $\mathbf{G}^{a}\left(t^{\prime},t_{0}\right)\propto\theta\left(t_{0}-t^{\prime}\right)$
for $t^{\prime}>t_{0}$. The second term $\mathcal{G}^{(2)}$ in the
first integral in Eq. (\ref{eq:long_G-tilde}) includes the convolution
integral $\left(\mathbf{\Sigma}^{\urcorner}\star\mathbf{G}^{\ulcorner}\right)_{\left(\bar{t},t_{0}\right)}$.
Using Eq. (\ref{eq:Left-GF-via-integral}) for the left Green's function,
we replace $\mathbf{G}^{\ulcorner}\left(\tau,t_{0+}\right)$ in the
convolution integral $\left(\mathbf{\Sigma}^{\urcorner}\star\mathbf{G}^{\ulcorner}\right)_{\left(\bar{t},t_{0+}\right)}$
with $\mathbf{G}^{M}\left(\tau,0^{+}\right)$. Using Eq. (\ref{eq:left-self-energy})
for the left self-energy and expanding the Matsubara Green's function
into the Fourier series (\ref{eq:Matsubara-GF-expansion}) and integrating
over $\tau$, we obtain:
\[
\left(\mathbf{\Sigma}^{\urcorner}\star\mathbf{G}^{\ulcorner}\right)_{\left(\bar{t},t_{0+}\right)}=\left(\sum_{\alpha}\mathbf{\Gamma_{\alpha}}e^{-i\psi_{\alpha}\left(\overline{t},t_{0}\right)}\right)\int\frac{d\omega^{\prime}}{2\pi}e^{-i\omega^{\prime}\left(\overline{t}-t_{0}\right)}\frac{i}{\beta}\sum_{q}\frac{\mathbf{G}^{M}\left(\omega_{q}\right)}{\omega_{q}-\omega^{\prime}+\mu}e^{\omega_{q}0^{+}}
\]
The sum over Matsubara frequencies is transformed using the well-known
formula
\begin{equation}
\frac{i}{\beta}\sum_{q}Q\left(\omega_{q}\right)e^{\omega_{q}0^{+}}=\int\frac{d\omega}{2\pi}f\left(\omega\right)\left[Q\left(\omega^{-}\right)-Q\left(\omega^{+}\right)\right]\label{eq:Matisumabara-sum-formula}
\end{equation}
where $\omega^{\pm}=\omega\pm i0$. Then the $\omega^{\prime}-$integration
is easily perfromed in the complex plane, and we obtain for the second
term in the first integral in Eq. (\ref{eq:long_G-tilde}):

\[
\mathcal{G}^{(2)}=\int\frac{d\omega}{2\pi}f\left(\omega-\mu\right)\underset{\alpha}{\sum}\mathbf{K}_{\alpha}\left(t_{1};\omega\right)\mathbf{\Gamma}_{\alpha}\mathbf{G}^{a}\left(\omega\right)
\]
where the $\mathbf{K}_{\alpha}$ matrix is given by Eq. (\ref{eq:K-integral})
and we have used the fact, following from the comparison of Eqs. (\ref{eq:Matsubara-GF-components})
and (\ref{eq:Advanced-GF-in-omega}), that $\mathbf{G}^{M}\left(\omega^{-}-\mu\right)\equiv\mathbf{G}^{a}\left(\omega\right)$.

The third contribution to the line integral coming from the term $\left(\mathbf{G}^{r}\cdot\mathbf{\Sigma}^{<}\right)_{\left(t_{1},\bar{t}\right)}$
in the second integral in Eq. (\ref{eq:long_G-tilde}) is obtained
immediately owing to simple expressions (\ref{eq:Retarded-GF-via-times})
and (\ref{eq:Lesser-self-energy}) for the retarded Green's function
and the lesser self-energy, respectively:
\[
\mathcal{G}^{(3)}=i\int\frac{d\omega}{2\pi}f\left(\omega-\mu\right)\underset{\alpha}{\sum}\mathbf{K}_{\alpha}\left(t_{1};\omega\right)\mathbf{\Gamma}_{\alpha}\mathbf{K}_{\alpha}^{\dagger}\left(t_{2};\omega\right)
\]
Finally, to calculate the last object $\mathcal{G}^{(4)}$ in the
expression (\ref{eq:long_G-tilde}), one needs the term $\left(\mathbf{G}^{\urcorner}\star\mathbf{\Sigma}^{\ulcorner}\right)_{\left(t_{1},\bar{t}\right)}$
which is first manipulated into the expression

\begin{equation}
\left(\mathbf{G}^{\urcorner}\star\mathbf{\Sigma}^{\ulcorner}\right)_{\left(t_{1},\bar{t}\right)}=e^{-i\mathbf{h}^{eff}\left(t_{1}-t_{0}\right)}\left[\left(\mathbf{G}^{M}\star\mathbf{\Sigma}^{\ulcorner}\right)_{\left(0^{-},\bar{t}\right)}-i\intop_{t_{0}}^{t_{1}}d\bar{t}'e^{i\mathbf{h}^{eff}\left(\bar{t}'-t_{0}\right)}\left(\mathbf{\Sigma}^{\urcorner}\star\mathbf{G}^{M}\star\mathbf{\Sigma}^{\ulcorner}\right)_{\left(\bar{t}',\bar{t}\right)}\right]\label{eq:2nd-int-interm1}
\end{equation}
using the formula (\ref{eq:Right-GF-via-integral}) for the right
Green's function. Above, in the second term we have a double convolution
integral along the imaginary track. However, it is straightforward
to show using explicit expressions (\ref{eq:right-self-energy}) and
(\ref{eq:left-self-energy}) for both self-energies and the expansion
(\ref{eq:Matsubara-GF-expansion}) for the Matsubara Green's function,
that this term is zero: 
\[
\left(\mathbf{\Sigma}^{\urcorner}\star\mathbf{G}^{M}\star\mathbf{\Sigma}^{\ulcorner}\right)_{\left(\bar{t}',\bar{t}\right)}=\frac{i}{\beta}\sum_{q}\left(\underset{\alpha}{\sum}\mathbf{\Gamma}_{\alpha}e^{-i\psi_{\alpha}\left(\bar{t}',t_{0}\right)}\right)\mathbf{G}^{M}\left(\omega_{q}\right)\left(\sum_{\alpha^{\prime}}\mathbf{\Gamma}_{\alpha'}e^{i\psi_{\alpha'}\left(\bar{t},t_{0}\right)}\right)
\]
\[
\times\int\frac{d\omega_{1}}{2\pi}\frac{e^{-i\omega_{1}\left(\bar{t}'-t_{0}\right)}}{\omega_{q}-\omega_{1}+\mu}\int\frac{d\omega_{2}}{2\pi}\frac{e^{i\omega_{2}\left(\bar{t}-t_{0}\right)}}{\omega_{q}-\omega_{2}+\mu}
\]
Indeed, the $\omega_{1}$ integral is only non-vanishing when $Im\left(\omega_{q}\right)<0$,
while the $\omega_{2}$ integral survives only when $Im\left(\omega_{q}\right)>0$.
We thus need only to calculate the following expression contained
in the first term in (\ref{eq:2nd-int-interm1}):

\[
\left(\mathbf{G}^{M}\star\mathbf{\Sigma}^{\ulcorner}\right)_{\left(0^{-},\bar{t}\right)}=\int\frac{d\omega}{2\pi}e^{i\omega\left(\bar{t}-t_{0}\right)}\frac{i}{\beta}\sum_{q}\frac{\mathbf{G}^{M}\left(\omega_{q}\right)}{\omega_{q}-\omega+\mu}e^{\omega_{q}0^{+}}\left(\underset{\alpha}{\sum}\mathbf{\Gamma}_{\alpha}e^{i\psi_{\alpha}\left(\bar{t},t_{0}\right)}\right)
\]
Once again, we perform the summation over the Matsubara frequencies
using formula (\ref{eq:Matisumabara-sum-formula}) and then perform
the $\omega$ integration in the complex plane, leading to the following
result for the final contribution to the line integral:

\[
\mathcal{G}^{(4)}=-\int\frac{d\omega}{2\pi}f\left(\omega-\mu\right)\mathbf{G}^{r}\left(\omega\right)\underset{\alpha}{\sum}\mathbf{\Gamma}_{\alpha}\mathbf{K}_{\alpha}^{\dagger}\left(t_{2};\omega\right)
\]

Summing up all four terms, and taking into account that 
\[
\mathbf{G}^{M}\left(0,0^{+}\right)=i\int\frac{d\omega}{2\pi}f\left(\omega-\mu\right)\underset{\alpha}{\sum}\mathbf{A}_{\alpha}\left(\omega\right)
\]
we obtain the result (\ref{eq:final-current}) given in the text.

\section*{Appendix C - Exact Expression for the Current Implementation}

When one substitutes (\ref{eq: dotKL}) and (\ref{eq: dotKR}) into
the formula (\ref{eq:final-current}), one can separate the resulting
formula into three terms, two of which decay due to factors of $e^{-\Gamma\left(t-t_{0}\right)/2}$
and $e^{-\Gamma\left(t-t_{0}\right)}$. These expressions are stated
here for completeness:

\begin{eqnarray*}
I_{L}^{\left(1\right)}\left(t\right) & = & e^{-\frac{\Gamma}{2}\left(t-t_{0}\right)}4\int\frac{d\omega}{2\pi}f\left(\omega-\mu\right)\Gamma_{L}\left[Im\left[e^{i\left(\omega+V_{L}-\varepsilon_{0}\right)\left(t-t_{0}\right)}e^{i\frac{A_{L}}{\Omega_{L}}\sin\left(\Omega_{L}t\right)}\right.\right.
\end{eqnarray*}

\[
\left.\times\left(\underset{n}{\sum}J_{n}\left(\frac{A_{L}}{\Omega_{L}}\right)G^{r}\left(\omega+V_{L}+n\Omega_{L}\right)e^{-in\Omega_{L}t_{0}}-G^{r}\left(\omega\right)e^{-i\frac{A_{L}}{\Omega_{L}}\sin\left(\Omega_{L}t_{0}\right)}\right)\right]
\]

\[
+Re\left[\Gamma_{L}\underset{n}{\sum}J_{n}\left(\frac{A_{L}}{\Omega_{L}}\right)G^{a}\left(\omega+V_{L}+n\Omega_{L}\right)e^{i\left(\omega+V_{L}-\varepsilon_{0}\right)\left(t-t_{0}\right)}e^{in\Omega_{L}t}\right.
\]

\begin{equation}
\left.\times\left(\underset{m}{\sum}J_{m}\left(\frac{A_{L}}{\Omega_{L}}\right)G^{r}\left(\omega+V_{L}+m\Omega_{L}\right)e^{-im\Omega_{L}t_{0}}-G^{r}\left(\omega\right)e^{i\frac{A_{L}}{\Omega_{L}}\sin\left(\Omega_{L}t_{0}\right)}\right)\right]\label{eq: current_gam2}
\end{equation}

\[
I^{\left(2\right)}\left(t\right)=e^{-\Gamma\left(t-t_{0}\right)}2\int\frac{d\omega}{2\pi}f\left(\omega-\mu\right)\Gamma_{L}^{2}\left[2Re\left[\underset{n}{\sum}J_{n}\left(\frac{A_{L}}{\Omega_{L}}\right)e^{i\frac{A_{L}}{\Omega_{L}}\sin\left(\Omega_{L}t_{0}\right)}e^{-in\Omega_{L}t_{0}}G^{r}\left(\omega+V_{L}+n\Omega_{L}\right)G^{a}\left(\omega\right)\right]\right.
\]

\textcolor{black}{
\[
\left.-\underset{n,m}{\sum}J_{n}\left(\frac{A_{L}}{\Omega_{L}}\right)J_{m}\left(\frac{A_{L}}{\Omega_{L}}\right)G^{r}\left(\omega+V_{L}+n\Omega_{L}\right)G^{a}\left(\omega+V_{L}+m\Omega_{L}\right)e^{i\Omega_{L}\left(m-n\right)t_{0}}-G^{r}\left(\omega\right)G^{a}\left(\omega\right)\right]
\]
}

The long-time behaviour of the current discussed in \ref{sec:Results}
is therefore given by the remaining term in the current formula:

\begin{eqnarray}
\underset{t_{0}\rightarrow-\infty}{\lim}I_{L}\left(t\right) & = & 2\int\frac{d\omega}{2\pi}f\left(\omega-\mu\right)\Gamma_{L}\left[2Im\left[\underset{n}{\sum}J_{n}\left(\frac{A_{L}}{\Omega_{L}}\right)e^{-i\frac{A_{L}}{\Omega_{L}}\left(\sin\left(\Omega_{L}t\right)-\sin\left(\Omega_{L}t_{0}\right)\right)}e^{in\Omega_{L}t}G^{a}\left(\omega+V_{L}+n\Omega_{L}\right)\right]\right.\label{eq: longtime}
\end{eqnarray}

\[
\left.-\underset{n,m}{\sum}J_{n}\left(\frac{A_{L}}{\Omega_{L}}\right)J_{m}\left(\frac{A_{L}}{\Omega_{L}}\right)e^{i\left(m-n\right)\Omega_{L}t}G^{r}\left(\omega+V_{L}+n\Omega_{L}\right)\Gamma_{L}G^{a}\left(\omega+V_{L}+m\Omega_{L}\right)-G^{r}\left(\omega\right)\Gamma_{R}G^{a}\left(\omega\right)\right]
\]

\begin{eqnarray*}
\\
\end{eqnarray*}


\begin{thebibliography}{10}

\bibitem{matsubara-PTP-1955}
T.~Matsubara.
\newblock {\em Prog. Theor. Phys.}, 14(4):351, 1955.

\bibitem{Martin-Schwinger-PR-1959}
Paul~C. Martin and Julian Schwinger.
\newblock {\em Phys. Rev.}, 115:1342, 1959.

\bibitem{konstantinov-perel-JETP-1961}
O.~V. Konstantinov and V.~I. Perel'.
\newblock {\em Sov. Phys. JETP}, 12(1):142, 1961.

\bibitem{KB1962}
L.~P. Kadanoff and G.~Baym.
\newblock {\em Quantum statistical mechanics: Green's function methods in
  equilibrium and nonequilibrium problems}.
\newblock New York: Benjamin, 1962.

\bibitem{Keldysh1964}
Leonid~V. Keldysh.
\newblock {\em Sov. Phys. JETP}, 20(4), 1964.

\bibitem{Langreth1976}
D.~C. Langreth.
\newblock {\em Linear and Non-Linear Response Theory with Applications},
  volume~17 of {\em NATO Advanced Studies Series B}, page~3.
\newblock Plenum, New York/London, 1976.

\bibitem{RammerSmith1986}
J.~Rammer and H.~Smith.
\newblock {\em Rev. Mod. Phys.}, 58:323, 1986.

\bibitem{RVLS2013}
G.~Stefanucci and R.~van Leeuwen.
\newblock {\em Nonequilibrium Many-Body Theory of Quantum Systems}.
\newblock Cambridge University Press, 2013.

\bibitem{KB1961}
G.~Baym and L.~P. Kadanoff.
\newblock {\em Phys. Rev.}, 124:287, 1961.

\bibitem{QD1997}
D.~L. Klein, R.~Roth, A.~Lim, P.~Alivisatos, and P.~L. McEuen.
\newblock {\em Nature}, 389(6652):699, 1997.

\bibitem{LANDAUER1957}
R.~Landauer.
\newblock {\em IBM J. Res. Dev.}, 1(3):223, 1957.

\bibitem{LANDAUER1970}
R.~Landauer.
\newblock {\em Phil. Mag.}, 21(172):863, 1970.

\bibitem{LandauerButtiker1982}
M.~B\"uttiker and R.~Landauer.
\newblock {\em Phys. Rev. Lett.}, 49:1739, 1982.

\bibitem{BUTTIKER1990}
M.~B\"uttiker.
\newblock {\em Phys. Rev. Lett.}, 65:2901, 1990.

\bibitem{BUTTIKER1992}
M.~B\"uttiker.
\newblock {\em Phys. Rev. B}, 46:12485, 1992.

\bibitem{diventra2001DFT}
M.~Di~Ventra and N.~D. Lang.
\newblock {\em Phys. Rev. B}, 65:045402, 2001.

\bibitem{diventra2001temp}
M.~Di~Ventra, S.-G. Kim, S.~T. Pantelides, and N.~D. Lang.
\newblock {\em Phys. Rev. Lett.}, 86:288, 2001.

\bibitem{vanruitenbeek1995}
J.~M. Krans, J.~M. Van~Ruitenbeek, V.~V. Fisun, I.~K. Yanson, and L.~J.
  De~Jongh.
\newblock {\em Nature}, 375(6534):767, 1995.

\bibitem{vanruitenbeek2003}
N.~Agrait, A.~L. Yeyati, , and J.~M.~Van Ruitenbeek.
\newblock {\em Phys. Rep.}, 377(2):81, 2003.

\bibitem{venkataraman2006}
L.~Venkataraman, J.~E. Klare, C.~Nuckolls, M.~S. Hybertsen, and M.~L.
  Steigerwald.
\newblock {\em Nature}, 442(7105):904, 2006.

\bibitem{blantbutt}
Y.~M. Blanter and B{\"u}ttiker. M.
\newblock {\em Phys. Rep.}, 336(1):1, 2000.

\bibitem{Burke2004}
P.~J. Burke.
\newblock {\em Solid State Electron.}, 48:1981, 2004.

\bibitem{Li2004}
S.~D. Li, Z.~Yu, S.-F. Yen, W.~C. Tang, and P.~J. Burke.
\newblock {\em Nano. Lett.}, 4(4):753, 2004.

\bibitem{Lin2009}
Y.~M. Lin, K.~A. Jenkins, A.~Valdes-Garcia, J.~P. Small, D.~B. Farmer, and
  P.~Avouris.
\newblock {\em Nano. Lett.}, 9(1):422, 2009.

\bibitem{MOSKALETS2012}
M.~Moskalets.
\newblock {\em Scattering Matrix Approach to Non-Stationary Quantum Transport}.
\newblock Imperial College Press, 2012.

\bibitem{CAROLI1}
C.~Caroli, R.~Combescot, P.~Nozieres, and D.~Saint-James.
\newblock {\em J. Phys. C: Solid State Phys.}, 4(8):916, 1971.

\bibitem{CAROLI2}
C.~Caroli, R.~Combescot, D.~Lederer, P.~Nozieres, and D.~Saint-James.
\newblock {\em J. Phys C: Solid State Phys.}, 4(16):2598, 1971.

\bibitem{JAUHO1993}
N.~S. Wingreen, A.-P. Jauho, and Y.~Meir.
\newblock {\em Phys. Rev. B}, 48:8487, Sep 1993.

\bibitem{JAUHO1994}
A.-P. Jauho, N.~S. Wingreen, and Y.~Meir.
\newblock {\em Phys. Rev. B}, 50:5528, 1994.

\bibitem{diventra}
M.~Di~Ventra.
\newblock {\em Electrical transport in nanoscale systems}.
\newblock Cambridge University Press, 2008.

\bibitem{CINI1980}
M.~Cini.
\newblock {\em Phys. Rev. B}, 22:5887, 1980.

\bibitem{StefALm2004}
G.~Stefanucci and C.-O. Almbladh.
\newblock {\em Phys. Rev. B}, 69:195318, 2004.

\bibitem{RVL2014}
R.~Tuovinen, E.~Perfetto, G.~Stefanucci, and R.~van Leeuwen.
\newblock {\em Phys. Rev. B}, 89:085131, 2014.

\bibitem{RVL2013}
Riku Tuovinen, Robert van Leeuwen, Enrico Perfetto, and Gianluca Stefanucci.
\newblock {\em J. Phys.: Conf. Ser.}, 427:012014, 2013.

\bibitem{RVLS2006}
N.~E. Dahlen, R.~van Leeuwen, and A.~Stan.
\newblock 35(1):324, 2006.

\bibitem{RVL2006}
R.~van Leeuwen, N.~E. Dahlen, G.~Stefanucci, C.-O. Almbladh, and U.~von Barth.
\newblock {\em Time-Dependent Density Functional Theory}, chapter Introduction
  to the Keldysh Formalism.
\newblock Springer, 2006.

\bibitem{RVL2009}
P.~My\"oh\"anen, A.~Stan, G.~Stefanucci, and R.~van Leeuwen.
\newblock {\em Phys. Rev. B}, 80:115107, 2009.

\bibitem{Yang2010}
S.-H. Ke, R.~Liu, W.~Yang, and H.~U. Baranger.
\newblock {\em J. Chem. Phys.}, 132(23), 2010.

\bibitem{Langreth1972}
D.~C. Langreth and J.~W. Wilkins.
\newblock {\em Phys. Rev. B}, 6(9):3189, 1972.

\bibitem{MRAMLK2014}
M.~Ridley, A.~MacKinnon, and L.~Kantorovich.
\newblock in preparation.

\bibitem{Danielewicz1984}
P.~Danielewicz.
\newblock {\em Ann. Phys.}, 152:239, 1984.

\bibitem{STEFANUCCI2014}
A.~Kartsev, C.~Verdozzi, and G.~Stefanucci.
\newblock {\em Eur. Phys. J. B}, 87(1):1, 2014.

\bibitem{RVLS2014}
S.~Latini, E.~Perfetto, A.-M. Uimonen, R.~van Leeuwen, and G.~Stefanucci.
\newblock {\em Phys. Rev. B}, 89:075306, 2014.

\bibitem{LandImry1999}
Y.~Imry and R.~Landauer.
\newblock {\em Rev. Mod. Phys.}, 71:S306, 1999.

\bibitem{DiVentraDFT2000}
M.~Di~Ventra, S.~T. Pantelides, and N.~D. Lang.
\newblock {\em Phys. Rev. Lett.}, 84:979, 2000.

\bibitem{HANGGI2006}
S.~Kohler, J.~Lehmann, and P.~H{\"a}nggi.
\newblock {\em Physics Reports}, 406(6):379, 2006.

\end{thebibliography}
\end{document}